\documentclass[AMA,STIX1COL]{SiM/WileyNJD-v2}

\usepackage{amsmath}
\usepackage{graphicx}
\usepackage[colorinlistoftodos]{todonotes}
\usepackage{booktabs}
\usepackage{float}
\usepackage{caption}
\usepackage{amsmath}
\usepackage{amssymb}
\usepackage{relsize}
\usepackage{moreverb}
\usepackage{ragged2e}
\usepackage{hhline}
\usepackage{xcolor} 
\usepackage{amssymb} 

\newcommand\BibTeX{{\rmfamily B\kern-.05em \textsc{i\kern-.025em b}\kern-.08em
T\kern-.1667em\lower.7ex\hbox{E}\kern-.125emX}}
\newcommand{\opt}{\text{opt}}
\newcommand{\HTE}{\text{HTE}}
\newcommand{\ATE}{\text{ATE}}
\newcommand{\BMI}{\text{BMI}}
\newcommand{\IFG}{\text{IFG}}

\articletype{Research Article}%

\received{<day> <Month>, <year>}
\revised{<day> <Month>, <year>}
\accepted{<day> <Month>, <year>}

\begin{document}

\title{Maximin optimal cluster randomized designs for assessing treatment effect heterogeneity}

\author[1,2]{Mary M. Ryan*}

\author[1,2]{Denise Esserman}

\author[1,2,3]{Fan Li}

\authormark{RYAN \textsc{et al}}

\address[1]{\orgdiv{Department of Biostatistics}, \orgname{Yale School of Public Health}, \orgaddress{\state{Connecticut}, \country{USA}}}

\address[2]{\orgdiv{Yale Center for Analytical Sciences}, \orgname{Yale School of Public Health}, \orgaddress{\state{Connecticut}, \country{USA}}}

\address[3]{\orgdiv{Center for Methods in Implementation and Prevention Science}, \orgname{Yale School of Public Health}, \orgaddress{\state{Connecticut}, \country{USA}}}

\corres{*Mary M. Ryan,\\
Department of Biostatistics,\\
Yale School of Public Health,\\
New Haven, Connecticut, USA\\
\email{mary.ryan@yale.edu}}


\abstract[Abstract]{Cluster randomized trials (CRTs) are studies where treatment is randomized at the cluster level but outcomes are typically collected at the individual level. When CRTs are employed in pragmatic settings, baseline population characteristics may moderate treatment effects, leading to what is known as heterogeneous treatment effects (HTEs). Pre-specified, hypothesis-driven HTE analyses in CRTs can enable an understanding of how interventions may impact subpopulation outcomes. While closed-form sample size formulas have recently been proposed, assuming known intracluster correlation coefficients (ICCs) for both the covariate and outcome, guidance on optimal cluster randomized designs to ensure maximum power with pre-specified HTE analyses has not yet been developed. We derive new design formulas to determine the cluster size and number of clusters to achieve the locally optimal design (LOD) that minimizes variance for estimating the HTE parameter given a budget constraint. Given the LODs are based on covariate and outcome-ICC values that are usually unknown, we further develop the maximin design for assessing HTE, identifying the combination of design resources that maximize the relative efficiency of the HTE analysis in the worst case scenario. In addition, given the analysis of the average treatment effect is often of primary interest, we also establish optimal designs to accommodate multiple objectives by combining considerations for studying both the average and heterogeneous treatment effects. We illustrate our methods using the context of the Kerala Diabetes Prevention Program CRT, and provide an R Shiny app to facilitate calculation of optimal designs under a wide range of design parameters.}


\jnlcitation{\cname{%
\author{M.M. Ryan}, 
\author{D. Esserman}, and 
\author{F. Li}} (\cyear{2023}), 
\ctitle{Maximin optimal cluster randomized designs for assessing treatment effect heterogeneity}, \cjournal{Statistics in Medicine}, \cvol{0000;00:00--00}.}

\maketitle

\footnotetext{\textbf{Abbreviations:} ATE, average treatment effect; CRT, cluster randomized trial; HTE, heterogeneous treatment effect; ICC, intracluster correlation coefficient; LOD, locally optimal design}

\section{Introduction}\label{s:intro}

Cluster randomized trials (CRTs) -- studies where treatment is randomized at the cluster or group level -- are gaining popularity in clinical medicine, public health and implementation science research. These designs are chosen for a variety of reasons such as the natural occurrence or grouping of the treatment clusters, treatment contamination prevention, or logistical constraints that would make individual randomization infeasible.\cite{murray_design_1998,hayes_cluster_2017} When CRTs are employed in pragmatic settings where identification of heterogeneous subpopulations is an important objective, diverse population characteristics, which may be key effect modifiers driving the variations in patient's response to interventions,  are often collected at baseline leading to what is known as heterogeneous treatment effects (HTEs). 

Whereas many {exploratory} HTE analyses are performed \emph{post-hoc} and represent essential steps for generating future hypotheses, confirmatory HTE analyses are often pre-specified, hypothesis-driven and can require more rigorous planning at the design stage. Although the power analysis of the treatment-by-covariate interaction test has been relatively well-studied in individually randomized trials,\cite{brookes_subgroup_2004,shieh_detecting_2009,greenland_tests_1983} related methods for power analysis in CRTs have only received recent attention with the goal to enable a rigorous understanding of how system-level innovations may differentially impact outcomes for important subpopulations. \cite{spybrook_power_2016,dong_power_2018,yang_sample_2020,tong_accounting_2021,li_designing_2022}

With a pre-specified effect modifier, Yang et al\cite{yang_sample_2020} developed an analytical sample size and power formula to test the treatment-by-covariate interaction, making it possible to power CRTs \textit{a priori} for confirmatory HTE analyses. Similar to designing conventional CRTs to study the average treatment effect, the intracluster correlation coefficient (ICC) of the outcome, or outcome-ICC, plays an essential role in determining the power and necessary sample size for the HTE test. In addition, the analytical formula of Yang et al\cite{yang_sample_2020} further requires knowledge of the covariate-ICC, or ICC of the effect modifier. The covariate-ICC can be characterized as the fraction of between-cluster covariate variation relative to the total or marginal variation of the covariate, and measures the degree of similarity of the effect modifier in the same cluster. Although the sample size formula for HTE has been previously characterized in CRTs, the optimal sample size, or equivalently, \emph{optimal design}, for testing HTE has not yet been investigated. In the CRT literature, the optimal design refers to the combination of number of clusters and cluster size that maximizes the power of the significance test, given a total budget for sampling and measuring clusters and individuals. As argued in van Breukelen and Candel,\cite{van_breukelen_efficient_2015} the identification of the optimal design can be of strong relevance from a cost-effectiveness standpoint; this has become an important consideration in implementation science studies as it allows more studies to be conducted with the same grand budget. 

To date, the identification of an optimal CRT design has been restricted to the objective of maximizing the average treatment effect. For example, Snijders and Bosker\cite{snijders_standard_1993} were the first to derive the optimal cluster size for two-level CRTs analyzing the average treatment effect for continuous outcomes with a linear mixed model in the absence of other covariates. Raudenbush\cite{raudenbush_statistical_1997} updated this derivation to account for the inclusion of covariates for increased precision, which also leads to the introduction of the concept of covariate-ICC in a different context. Extensions to three-level CRTs,\cite{moerbeek_design_2000} logistic regression models with\cite{moerbeek_optimal_2005} and without\cite{moerbeek_optimal_2001} covariates, unequal costs between study arms,\cite{liu_statistical_2003} and multiple treatment effects collected at different levels\cite{moerbeek_optimal_2020} subsequently followed. For ease of reference, we provide a summary of existing optimal design methods for CRTs in Table \ref{tab:LODlit}. These approaches all suggest that the optimal design critically depends on the outcome-ICC, which drives the precision of the average treatment effect estimator. This means that an optimal CRT design derived under one outcome-ICC estimate will likely not be optimal under a different value; thus, such designs are only \emph{locally optimal}.

\begin{table}
\caption{\label{tab:LODlit}Brief summary of existing literature on locally optimal designs for cluster randomized trials that study the average treatment effect.}
\centering
{\RaggedRight
\begin{tabular}{l p{0.1\linewidth}p{0.1\linewidth}p{0.1\linewidth}p{0.07\linewidth}p{0.27\linewidth}}
\toprule
Reference & \multicolumn{2}{c}{CRT Design Type}  & \multicolumn{2}{c}{Outcome} & Feature\\
\cmidrule(lr){2-3} \cmidrule(lr){4-5}
& Two-level & Three-level & Continuous & Binary&\\
\midrule
Snijders \& Bosker (1993)\cite{snijders_standard_1993} & \checkmark & - & \checkmark & - & Introduces optimal design to CRTs\\
\midrule
Raudenbush (1997)\cite{raudenbush_statistical_1997} & \checkmark & - & \checkmark & - & Optimal design conditional on covariate\\
\midrule
Moerbeek et al (2000)\cite{moerbeek_design_2000} & \checkmark & \checkmark & \checkmark & - & Optimal designs for three-level CRTs\\
\midrule
Moerbeek et al (2001a)\cite{moerbeek_optimal_2001-1} & \checkmark & - & - & \checkmark & Optimal designs and randomization for CRTs using logistic models\\
\midrule
Moerbeek et al (2001b)\cite{moerbeek_optimal_2001} & \checkmark & - & \checkmark & - & Introduces D- and L-optimality criteria\\
\midrule
Liu (2003)\cite{liu_statistical_2003} & \checkmark & \checkmark & \checkmark & - & Optimal unequal allocation design for CRTs with unequal costs per randomization unit\\
\midrule
Moerbeek \& Maas (2005)\cite{moerbeek_optimal_2005} & \checkmark & - & - & \checkmark & Optimal designs for multilevel logistic model with covariates\\
\midrule
Moerbeek (2020)\cite{moerbeek_optimal_2020} & \checkmark & - & \checkmark & - & Optimal designs for multiple treatment effects\\
\bottomrule
\end{tabular}}
\end{table}

While reporting of the outcome-ICC is becoming more commonplace among CRTs, it can still be difficult to predict at the design stage and misspecification can severely impact sample size and power calculations. To mitigate this issue, van Breukelen and Candel\cite{van_breukelen_efficient_2015} introduced maximin designs for CRTs investigating the average treatment effect. Maximin design procedures find the most efficient design with respect to a budget constraint for a range of outcome-ICC values, meaning the design that maximizes power given a hard budget or minimizes budget given a power threshold in the worst case outcome-ICC scenarios. Liu et al\cite{liu_optimal_2019} extended this work to the setting of three-level CRTs.

All current locally optimal and maximin design methods for CRTs are specifically developed for assessing the average treatment effect; no attempts have yet been made to derive optimal procedures for assessing HTE which, as shown by Yang et al,\cite{yang_sample_2020} would critically depend on both outcome- and covariate-ICCs. In addition, to the best of our knowledge, the covariate-ICC is not standard in trial reporting and reliable information on reasonable ranges may be less available than for the outcome-ICC,\cite{korevaar_intra-cluster_2021} making its elucidation in CRT design procedures difficult. Thus, developing a maximin design procedure for testing HTE, by considering a range for the covariate-ICC with a fixed outcome-ICC, or considering a range for both the outcome- and covariate-ICC, may prove essential to designing CRTs adequately powered for HTE and answering pre-specified questions involving diverse subpopulations. This points to the central focus of this paper.

In addition, it is rare for testing of HTE hypotheses to be the sole aim of a study. Often, the average or main treatment effect is also of interest -- if not the primary interest -- and the sample sizes required to properly power each set of analyses may not align. It then becomes a question of how to strike a balance between these study objectives. Very little research has been conducted in this area for CRTs. Moerbeek\cite{moerbeek_optimal_2020} developed a multiple-objective optimal design procedure for CRTs when both individual- and cluster-level outcomes are of interest but only when other design parameters are fixed, creating a multiple-objective locally optimal design. Such procedures have not been extended to the maximin design space. To fill this gap, we will also extend the HTE optimal design procedures in the manner of Moerbeek\cite{moerbeek_optimal_2020} to balance considerations for both the heterogeneous and average treatment effect objectives.

The remainder of this article is organized as follows. In Section \ref{s:model}, we introduce the linear mixed analysis of covariance model with a pre-specified effect modifier and review the main result in Yang et al.\cite{yang_sample_2020} In Section \ref{s:sood}, we develop a closed-form solution for the locally optimal CRT design for assessing HTE with a pre-specified effect modifier, as well as a maximin design procedure that accommodates uncertainties in the covariate-ICC and outcome-ICC. In Section \ref{s:mood}, we expand the results of Section \ref{s:sood} to arrive at optimal designs when the objective function incorporates considerations on both the HTE analysis and the average treatment effect analysis, leading to the multiple-objective optimal designs. In Section \ref{s:power} we briefly discuss power considerations in practice. In Section \ref{s:dataApp} we use data from the Kerala Diabetes Prevention Program study (K-DPP)\cite{thankappan_peer-support_2018} to illustrate the proposed new optimal design procedures and determine the number of clusters and cluster sizes required to maximize power under a fixed grand budget. Finally, in Section \ref{s:discuss} we discuss the results and possible future work in this area. To facilitate the exploration of optimal designs for assessing HTE in a wider range of practical scenarios, we also provide a free R shiny application to implement the proposed procedure at:  \url{https://mary-ryan.shinyapps.io/HTE-MMD-app/}.

\section{Statistical model}\label{s:model}

Before we develop the optimal designs to assess HTE in CRTs, we first introduce the linear mixed analysis of covariance model, as well as review the existing sample size formulas developed in Yang et al.\cite{yang_sample_2020} We consider parallel CRTs with two arms. Let $Y_{ij}$ be a continuous outcome for the $j$th individual ($j = 1, \dots, m)$ in the $i$th cluster ($i = 1, \dots, n)$; we assume equal cluster sizes following the convention of deriving optimal designs. When we are solely interested in evaluating the average treatment effect, it is common to analyze the individual-level outcomes using a linear mixed effects model similar to the one outlined below:\cite{turner_review_2017}
\begin{equation}\label{eq:noInteract}
    Y_{ij} = \alpha_1+ \alpha_2 W_i + \psi_i + \xi_{ij},
\end{equation}
where $W_i$ is the binary treatment indicator ($W_i=1$ if cluster \textit{i} is assigned to intervention and $W_i=0$ otherwise), $\psi_i \sim \mathcal{N}(0, \sigma^2_{\psi})$ is the random cluster effect accounting for the outcome-ICC, and $\xi_{ij} \sim \mathcal{N}(0, \sigma^2_{\xi})$ is the residual error, independent of the random cluster effect. In this unadjusted regression model, $\alpha_1$ represents the mean of the outcome under the control condition and $\alpha_2$ represents the average treatment effect without adjusting for covariates.

A primary goal in pragmatic CRTs is to evaluate interventions in settings similar to those observed in the real world, i.e., settings with realistic population diversity and heterogeneity. The investigators therefore may be interested in testing for possible treatment effect heterogeneity with respect to a pre-specified effect modifier. To introduce the linear mixed model accounting for an effect modifier, we assume $X_{ij}$ is a continuous or binary univariate covariate that may moderate the treatment effect. We consider the effect modifier to be measured either at the individual-level or cluster-level; in the latter case, we can simply replace $X_{ij}$ with $X_i$ as all individuals in the same cluster will have the same value of the effect modifier when it is measured at the cluster level. For simplicity, we also assume the effect modifier to be univariate and will discuss possible extensions to multivariate effect modifiers in Section \ref{s:discuss}. With $X_{ij}$, model (\ref{eq:noInteract}) can be expanded as:
\begin{equation}\label{eq:interact}
    Y_{ij} = \beta_1 + \beta_2 W_i + \beta_3 X_{ij} + \beta_4 X_{ij} W_i + \gamma_i + \epsilon_{ij},
\end{equation}
where $X_{ij} W_i$ is the interaction between treatment and covariate, $\gamma_i \sim \mathcal{N}(0, \sigma^2_{\gamma})$ is the random cluster effect, and $\epsilon_{ij} \sim \mathcal{N}(0, \epsilon^2_{\epsilon})$ is the residual error, independent of $\gamma_i$. Of note, we have not considered a random slope for the effect modifier, such that there are no additional cluster-by-covariate interactions. In this analysis of covariance type model, $\beta_1$ is the mean of the outcome under the control condition when $X_{ij}=0$, $\beta_2$ is the treatment effect when $X_{ij}=0$ (or the average treatment effect when $X_{ij}$ is mean-centered at 0), and $\beta_3$ and $\beta_4$ are regression coefficients for the covariate and the interaction terms, respectively. In particular, the magnitude of $\beta_4$ can quantify the degree of treatment effect heterogeneity regarding the effect modifier, and can be considered as a basis for testing for HTE in CRTs. Further, if the covariate is mean-centered, $\beta_2$ represents the average treatment effect parameter under model \eqref{eq:interact}.\cite{tong_accounting_2021} Mean-centering the covariates, however, does not affect the interpretation of the interaction parameter $\beta_4$.

While sample size considerations (and subsequently optimal designs) based on the unadjusted linear mixed model \eqref{eq:noInteract} have been relatively well studied, sample size considerations based on the adjusted linear mixed model \eqref{eq:interact} have only been recently examined for applications to CRTs. Specifically, for the purpose of testing for HTE, Yang et al\cite{yang_sample_2020} showed that the variance of the maximum likelihood estimator $\hat{\beta}_4$, which we denote $\sigma^2_{\HTE}$, is:
\begin{equation}\label{eq:sHTE}
    \sigma^2_{\HTE} = \frac{\sigma^2_{y|x}(1-\rho_{y|x}) \left\{1+(m-1)\rho_{y|x}\right\}}{nm\sigma^2_w \sigma^2_x \left\{1+(m-2)\rho_{y|x} - (m-1)\rho_x\rho_{y|x}\right\}},
\end{equation}
where $m$ is the common cluster size, $n$ is the total number of clusters, $\sigma^2_{y|x}=\sigma_\gamma^2+\sigma_\epsilon^2$ is the total variance of $Y_{ij}$ adjusted for $X_{ij}$, $\sigma^2_x$ is the marginal variance of the covariate $X_{ij}$, and $\sigma^2_w = E(W_i)\{1-E(W_i)\}$ quantifies the variation in treatment assignment. Importantly, expression \eqref{eq:sHTE} also features two key intracluster correlation coefficients: $\rho_{y|x} = {\sigma^2_{\gamma}}/{\sigma^2_{y|x}}$ represents the outcome-ICC adjusted for $X_{ij}$, and $\rho_x$ represents the covariate-ICC; the latter concept can be defined as the fraction of between-cluster covariate variation relative to the total or marginal variation of the covariate, $\sigma_x^2$, and measures the degree of similarity of the effect modifier in the same cluster.\cite{raudenbush_statistical_1997}

Finally, Tong et al\cite{tong_accounting_2021} showed that when the cluster sizes are equal and the covariate is mean-centered (and assumed to be uncorrelated with the treatment variable in large samples), the variance of the covariate-adjusted average treatment effect estimator $\hat{\beta}_2$, which we denote $\sigma^2_{\ATE}$, is:
\begin{equation}\label{eq:sATE}
    \sigma^2_{\ATE} = \frac{\sigma^2_{y|x}\left\{1+(m-1)\rho_{y|x}\right\}}{nm\sigma^2_w},
\end{equation}
where $\left\{1+(m-1)\rho_{y|x}\right\}$ is commonly referred to as the design effect in CRTs. For subsequent purposes, we can also write $\sigma^2_{\HTE}$ in terms of $\sigma^2_{\ATE}$ with a multiplication factor as:
$$\sigma^2_{\HTE} = \sigma^2_{\ATE} \times \frac{(1-\rho_{y|x})}{\sigma^2_x\left\{1+(m-2)\rho_{y|x} - (m-1)\rho_x\rho_{y|x}\right\}}.$$
In the special case with a cluster-level covariate and $\rho_x=1$ by definition, we obtain $\sigma^2_{\HTE} = \sigma^2_{\ATE}/\sigma_x^2$. Note that in this case, the ratio $\sigma^2_{\HTE}/\sigma^2_{\ATE}$ does not depends on the  number of clusters $n$ nor the cluster size $m$, such that the optimal design would be the same for studying the average or heterogeneous treatment effects. On the other hand, in the special case where the individual-level covariate randomly varies both within and between clusters (such that the extra between-cluster variation is $0$), or $\rho_x= 0$, then the ratio $\sigma^2_{\HTE}/\sigma^2_{\ATE}$ decreases from $1/\sigma^2_x$ for $\rho_{y|x} = 0$ to $0$ for $\rho_{y|x}=1$.

\section{Optimal Designs for Assessing Treatment Effect Heterogeneity}\label{s:sood}

Determining an efficient CRT study design is rarely a simple task due to the confluence of enrolling both clusters and individuals, the uncertainty in design parameters, as well as budget restrictions. Here, we refer to a CRT design as the combinations of the total number of clusters $n$ and cluster size $m$. Designs are considered optimal if they minimize the variance of the estimator of interest given a fixed budget constraint, or if they minimize costs given a fixed level of precision; we will focus on the case where the budget constraint is fixed. We suppose we have a total budget $B$ to spend on our study. Assuming inclusion of each cluster in the study costs $c$ and inclusion of each individual subject within a cluster costs $s$, we can divide our total budget into the cost attributable to cluster and subject inclusion:
\begin{equation}\label{eq:budget}
    B=cn + smn = n(c + sm)
\end{equation}
In the special case where $c=0$ and $s=1$, equation \eqref{eq:budget} returns the traditional total sample size constraint: $B=nm$. We note that optimal CRT designs for estimating the average treatment effect have already been investigated extensively in the literature (see Table \ref{tab:LODlit}); thus, in what follows we will primarily focus on optimal designs for testing the HTE. First we will derive a closed-form solution for the locally optimal design (LOD) for testing the HTE, which relies on exact specification of ICC parameters. Then we will develop a maximin design procedure that is optimal over a range of outcome- and covariate-ICC values specified in the design stage.

\subsection{Locally optimal design}\label{ss:solod}
A single-objective LOD is one in which the highest efficiency or smallest variance is achieved for a single objective or estimator on a known set of parameters, given a budget constraint such as \eqref{eq:budget}. The single-objective optimal design for the HTE would be the one where, for known values of ($\rho_{y|x}$,$\rho_x$), $\sigma^2_{\HTE}$ is minimized. To achieve this, we rearrange the cost function \eqref{eq:budget} for $n$ and substitute this into variance equation \eqref{eq:sHTE}:
\begin{align}
\begin{split}\label{eq:sHTE2}
    \sigma^2_{\HTE} &\propto \frac{c+sm}{Bm}\times \frac{(1-\rho_{y|x})\left\{1+(m-1)\rho_{y|x}\right\}}{\left\{1+(m-2)\rho_{y|x} - (m-1)\rho_x\rho_{y|x}\right\}}\\
    &= \frac{s(1-\rho_{y|x})}{B}\times \frac{(k+m)\left\{1+(m-1)\rho_{y|x}\right\}}{m\left\{1+(m-2)\rho_{y|x} - (m-1)\rho_x\rho_{y|x}\right\}},
\end{split}
\end{align}
where the proportionality constant is $\sigma^2_{y|x}/(\sigma^2_w\sigma^2_x)$ and $k=c/s>0$ is the cluster-to-individual cost ratio. While $k$ technically need only be greater than $0$, the only instances where it would be less than $1$ would be special circumstances where individual-level costs might include very expensive individual data collection procedures or interventions (e.g., Magnetic resonance imaging (MRI); positron emission tomography (PET) scan).

Minimizing the above with respect to $m$, we obtain the closed-form LOD for testing the HTE:
\begin{align*}
\begin{split}
    m_{\opt} &= \frac{(1-\rho_{y|x})(1-\rho_x) + \sqrt{\rho^{-1}_{y|x}k^{-1}(1-\rho_{y|x})(\rho_x-\rho_{y|x})\left\{1-(k+2)\rho_{y|x}+(k+1)\rho_x\rho_{y|x}\right\}}}{k^{-1}(\rho_x-\rho_{y|x}) -\rho_{y|x}(1-\rho_x)},\\
    n_{\opt} &= \frac{B}{c+sm_{opt}},
\end{split}
\end{align*}
meaning that the design with the highest precision to test the HTE for a given budget and fixed ICC values $\rho_{y|x}$ and $\rho_x$ is one where there are a total of $n_{\opt}$ clusters, each of size $m_{\opt}$. We note that the optimal cluster size relies on budget constraint \eqref{eq:budget} only through the cost ratio, and does not further depend on the size of the total budget nor the precise per-unit cost of clusters or individuals.

The above closed-form LOD includes some explicit conditions on design parameters. To elaborate, in order for the optimal cluster size $m_{\opt}$ to be real and greater than $1$, there is an implied plausible range for the covariate-ICC, $\rho_x$. It can be shown that the above solution for optimal LOD is achieved when (assuming $k>0$) 
\begin{align*}
  \frac{\rho_{y|x}(k+1)}{\rho_{y|x}k+1} < \rho_x \le 1,~~~~\text{and}~~~~
  0 \le \rho_{y|x} < 1,
\end{align*}
We note that while $\rho_{y|x} \in [0,1)$, it rarely exceeds $0.2$.\cite{van_breukelen_efficient_2015} In addition, if covariate $X_{ij}$ is a good prognostic variable, its inclusion in model \eqref{eq:interact} can sometimes drive $\rho_{y|x}$ toward $0$ (due to explained variation, such as when $\rho_x$ is large), increasing the acceptable range for $\rho_x$. When $\rho_x$ is outside this valid range, such as when $\rho_x$ is close to $0$ and $\rho_{y|x}$ is relatively far away from $0$, we have observed in numerical evaluations that $\sigma^2_{\HTE}$ generally decreases as $m\rightarrow \infty$ and $n$ is decreased to remain within the budget constraint. In these scenarios, it would be reasonable to set $m_{\opt}$ to a maximum determined \emph{a priori}. A very large maximum would, under budget constraints, encourage a very small number of clusters. Since CRTs lose their utility with respect to individually randomized trials when designed with an extremely small number of clusters, we also want to \emph{a priori} specify a minimum for $n_{\opt}$, which we define as $\underline{n}$. We can then use this lower bound for the number of clusters and budget constraint \eqref{eq:budget} to define a maximum cluster size for $m_{\opt}$, given by $\overline{m}=(B/\underline{n}-c)/s$. This maximum can also be utilized even when $\rho_x$ is in the valid range but the unrestricted LOD calls for an $m_{\opt}$ that would drive $n$ below the minimum $\underline{n}$. To unify the above practical considerations, we propose a conditional LOD in Proposition \ref{PROP:SOLOD}.

\begin{proposition}\label{PROP:SOLOD}
\textit{Given a fixed budget constraint, a minimum number of clusters, an outcome-ICC, and a covariate-ICC, the locally optimal design for a cluster randomized trial that minimizes $\sigma^2_{\HTE}$ is given by:}
\begin{align*}
m_{\opt}=\frac{(1-\rho_{y|x})(1-\rho_x)+\sqrt{\rho^{-1}_{y|x} k^{-1} (1 - \rho_{y|x})(\rho_x - \rho_{y|x})\left\{1-(k+2)\rho_{y|x} + (k+1)\rho_x\rho_{y|x}\right\}}}{k^{-1}(\rho_x - \rho_{y|x}) - \rho_{y|x} (1 - \rho_x)}, 
\end{align*}
\textit{under the condition that}
\begin{align}\label{eq:cond}
\frac{\rho_{y|x}(k+1)}{\rho_{y|x}k+1} < \rho_x \le 1,~~~\text{and}~~~m_{\opt} \le \frac{B/\underline{n} - c}{s}.
\end{align}
\textit{If condition \eqref{eq:cond} is not satisfied, then we set}
\begin{align*}
m_{\opt} = \frac{B/\underline{n} - c}{s}.
\end{align*}
\textit{In either case, the optimal number of clusters is given by}
\begin{align*}
n_{\opt} = \frac{B}{c+sm_{\opt}}.
\end{align*}

\begin{proof}
See Appendix \ref{soLODProof}.
\end{proof}
\end{proposition}

As a concrete illustration, Table \ref{tab:soLOD} shows examples of LODs calculated via Proposition \ref{PROP:SOLOD} for combinations of known ICC values. In Table \ref{tab:soLOD}, we assume $B=100,000$, cost ratios of $k=10$ ($c=500$, $s=50$) and $k=20$ ($c=2,000$, $s=100$), and a minimum of $\underline{n}=6$ clusters. For the purpose of illustrating the power of each design, we select the standardized HTE effect size, defined by $\beta_4\sigma_x/\sigma_{y|x}=0.2$ and set $\sigma^2_{y|x} = \sigma^2_x = 1$. This standardized effect size is interpreted as the change in treatment effect (per standard deviation unit of the outcome) due to one standard deviation unit change in the effect modifier. We see that, for a fixed value of $\rho_{y|x}$, the optimal design shifts from a few large clusters to many small clusters as $\rho_x$ increases; this also results in a reduction in power. This pattern is consistent with the idea that as $\rho_x$ increases, it becomes more akin to a cluster-level covariate, which would make the number of clusters more important for estimating the HTE parameter, $\beta_4$. On the other hand, as $\rho_{y|x}$ increases, power becomes more sensitive to changes in $\rho_x$, confirming results observed by Yang et al\cite{yang_sample_2020} in fixed, non-optimal designs. We also see that if $\rho_x$ is held constant and is within its valid range, the optimal design generally shifts from a few large clusters to many small clusters as $\rho_{y|x}$ increases. However, we may see $m_{\opt}$ abruptly ``jump'' up when $\rho_x$ is near the lower bound of its valid range. For example, when $k=10$, $\rho_{y|x}=0.1$ and $\rho_x=0.75$, $m_{\opt}=20$ and the lower bound for $\rho_x=0.55$; when $\rho_{y|x}$ increases to $0.2$ and $\rho_x$ is kept fixed at $0.75$, though, the lower bound for $\rho_x$ increases to $0.733$ and $m_{\opt}$ ``jumps'' to $86$. Finally, we observe that as $\rho_{y|x}$ increases, so does the frequency with which $\rho_x$ is outside its valid range, forcing $m_{\opt}$ to take on the maximum cluster size value, $\overline{m}$, more frequently. To examine a wider range of ICC parameter values, the LOD for assessing HTE can also be implemented via a free web application at \url{https://mary-ryan.shinyapps.io/HTE-MMD-app/}.

Finally, in the special case where we are interested in testing HTE with respect to a cluster-level effect modifier (i.e., $\rho_x=1$), the optimal design simplifies to:
\begin{align*}
    m_{\opt} &= \frac{\sqrt{\rho^{-1}_{y|x} k^{-1} (1 - \rho_{y|x})}}{k^{-1}} = \sqrt{\frac{(1-\rho_{y|x})}{\rho_{y|x}}\times k} = \sqrt{\frac{\theta c}{s}},\\
n_{\opt} &=\frac{B}{\sqrt{\theta s c}+c},~~~~~~
\theta = \frac{1-\rho_{y|x}}{\rho_{y|x}}.
\end{align*}
This optimal design shares the same form with the optimal CRT design for testing the average treatment effect developed in Raudenbush,\cite{raudenbush_statistical_1997} Moerbeeek et al,\cite{moerbeek_design_2000} and van Breukelen and Candel.\cite{van_breukelen_efficient_2015} This is expected because the variance for the interaction parameter in linear mixed model \eqref{eq:interact} includes the same design effect as appears in the variance for the average treatment effect in CRTs (also see Section \ref{s:model}).

\begin{table}
\caption{\label{tab:soLOD}Locally optimal design with cluster size ($m$),  number of clusters ($n$), and power to detect a standardized HTE effect size of 0.2 for known outcome-ICC ($\rho_{y|x}$) and covariate-ICC ($\rho_x$) values assuming a total budget $B=100,000$, cost ratios of $k=10$ ($c=500$, $s=50$) and $k=20$ ($c=2,000$, $s=100$), and $\sigma^2_{y|x} = \sigma^2_x = 1$. Bold values indicate instances where $\rho_x$ is outside the valid range and the maximum cluster size (minimum six clusters) was used as optimal.}
\centering
\begin{tabular}{ll | lll | lll}
\toprule
&& \multicolumn{3}{c |}{Cost ratio $k=10$} & \multicolumn{3}{c}{Cost ratio $k=20$}\\ 
$\rho_{y|x}$ & $\rho_x$ & $m$ & $n$ & Power& $m$ & $n$ & Power\\
\midrule
0.005 & 0.1 & \textbf{323} & \textbf{6} & \textbf{0.990} & \textbf{146} & \textbf{6} & \textbf{0.826}\\
 & 0.2 & 175 & 10 & 0.979 & \textbf{146} & \textbf{6} & \textbf{0.809}\\
 & 0.5 & 76 & 23 & 0.973 & 119 & 7 & 0.741\\
 & 0.75 & 55 & 30 & 0.961 & 81 & 9 & 0.668\\
 & 1 & 44 & 37 & 0.955 & 63 & 12 & 0.671\\
\midrule
0.05 & 0.1 & \textbf{323} & \textbf{6} & \textbf{0.990} & \textbf{146} & \textbf{6} & \textbf{0.824}\\
 & 0.2 & \textbf{323} & \textbf{6} & \textbf{0.982} & \textbf{146} & \textbf{6} & \textbf{0.784}\\
 & 0.5 & 61 & 28 & 0.913 & \textbf{146} & \textbf{6} & \textbf{0.618}\\
 & 0.75 & 22 & 62 & 0.830 & 40 & 16 & 0.441\\
 & 1 & 13 & 86 & 0.753 & 19 & 25 & 0.352\\
\midrule
0.1 & 0.1 & \textbf{323} & \textbf{6} & \textbf{0.993} & \textbf{146} & \textbf{6} & \textbf{0.841}\\
 & 0.2 & \textbf{323} & \textbf{6} & \textbf{0.986} & \textbf{146} & \textbf{6} & \textbf{0.800}\\
 & 0.5 & \textbf{323} & \textbf{6} & \textbf{0.913} & \textbf{146} & \textbf{6} & \textbf{0.619}\\
 & 0.75 & 20 & 66 & 0.751 & 74 & 10 & 0.376\\
 & 1 & 9 & 105 & 0.630 & 13 & 30 & 0.265\\
\midrule
0.2 & 0.1 & \textbf{323} & \textbf{6} & \textbf{0.997} & \textbf{146} & \textbf{6} & \textbf{0.880}\\
 & 0.2 & \textbf{323} & \textbf{6} & \textbf{0.993} & \textbf{146} & \textbf{6} & \textbf{0.841}\\
 & 0.5 & \textbf{323} & \textbf{6} & \textbf{0.938} & \textbf{146} & \textbf{6} & \textbf{0.657}\\
 & 0.75 & 86 & 20 & 0.690 & \textbf{146} & \textbf{6} & \textbf{0.403}\\
 & 1 & 6 & 125 & 0.491 & 8 & 35 & 0.189\\
\bottomrule
\end{tabular}
\end{table}

\subsection{Maximin design}\label{ss:sommd}

Section \ref{ss:solod} illustrated how the optimal design that minimizes $\sigma^2_{\HTE}$ within a budget constraint varies with the outcome-ICC and covariate-ICC. While reporting of the outcome-ICC is recommended practice for parallel CRTs\cite{campbell_consort_2004,campbell_consort_2012} and becoming increasingly commonplace, reporting of covariate-ICC is currently uncommon. Thus there is likely to be substantial uncertainty around these values at the design stage, and misspecification of ICC values can result in inaccurate sample size estimates and lead to either over- or under-powered trials.

To address this potential limitation for designing studies interested in assessing the average treatment effect, van Breukelen and Candel\cite{van_breukelen_efficient_2015} introduced a maximin CRT design procedure. Through a search process, this procedure identifies a design that is optimal for a particular outcome-ICC value while getting as close as possible to the maximum relative efficiency (RE) for the other values in a given plausible range. We consider a similar procedure but now focus on the assessment of HTE in CRTs. Specifically, we define RE for assessing the HTE as:
\begin{equation}
    \text{RE}_{\HTE} = \frac{\sigma^{2*}_{\HTE}}{\sigma^2_{\HTE}},
\end{equation}
where $\sigma^{2*}_{\HTE}$ is the variance of the HTE parameter estimator under the LOD from Section \ref{ss:solod}. Based on this RE expression, we extended the maximin design procedure to accommodate uncertainty in both the outcome-ICC and the covariate-ICC for assessing the HTE in CRT. This maximin design procedure for testing HTE is summarized in Algorithm \ref{algo:soMMD}.
\begin{algorithm}
\caption{Maximin design procedure for assessing HTE in CRTs}\label{algo:soMMD}
\begin{algorithmic}[1]
    \State Define the discrete parameter space for ($\rho_{y|x}$, $\rho_x$) and design space for $\left(m, n(m) \right)$;
    \State For each ($\rho_{y|x}$, $\rho_x$) parameter value combination, compute the LOD for the HTE objective using Proposition \ref{PROP:SOLOD}. Then compute the RE for each $\left(m, n(m) \right)$ design value combination compared with the LOD at the parameter value pair by taking the ratio of the variances;
    \State For each $\left(m, n(m) \right)$ design value combination, identify the ($\rho_{y|x}$, $\rho_x$) parameter value combination with the smallest RE;
    \State Among the smallest REs, choose the $\left(m, n(m) \right)$ design value combination with the largest RE. This returns the maximin optimal design for assessing HTE in CRTs. 
\end{algorithmic}
\end{algorithm}

Of note, the maximin design in Algorithm \ref{algo:soMMD} is not an exhaustive search over every ($m$, $n$) combination; instead $n$ is determined as a function of $m$ via ${B}/(c+sm)$ (also see Proposition \ref{PROP:SOLOD}). It also need not be an exhaustive search over the ICC parameter space; similar to observations made by van Breukelen and Candel,\cite{van_breukelen_efficient_2015} we observe that the maximin design for assessing HTE is often found at the intersection of two out of four potential RE curves defined by the boundaries of the ICC parameter ranges, when the design space is relatively broad in $m$: ($\underline{\rho_{y|x}},\underline{\rho_x}$), ($\underline{\rho_{y|x}},\overline{\rho_x}$), ($\overline{\rho_{y|x}},\underline{\rho_x}$), ($\overline{\rho_{y|x}},\overline{\rho_x}$), where $\underline{\rho_{y|x}}$ and $\overline{\rho_{y|x}}$ refer to the minimum and maximum values of the outcome-ICC in the specified parameter space, and $\underline{\rho_{x}}$ and $\overline{\rho_{x}}$ refer to the minimum and maximum values of the covariate-ICC in the specified parameter space, respectively.

There are several cases where the maximin design will not be found at an intersection between these scenarios, but at the maximum value of $m$ in the design space. First, a larger cost-ratio $k$ will flatten RE curves for all ICC scenarios such that LODs are found at larger $m$ to offset the relatively increased cost of additional clusters; thus, intersections between scenarios will occur at larger values of $m$ and if the design space is restricted, $\overline{m}$ may be smaller than this potential intersection point. Second, smaller maximum values of the outcome- and covariate-ICCs will flatten RE curves for ICC scenarios involving the maximums, and the LODs for these scenarios are found at larger values of $m$ due to a lower degree of clustering; the maximum of the covariate-ICC is usually more influential for this than the outcome-ICC. If the design space does not extend to these regions, the maximin design will be found at the maximum value of $m$ in the design space.

As an illustration, Figure \ref{fig:soMMD} shows two examples of maximin designs for assessing HTE where $\rho_{y|x} \in [0.005, 0.2]$, $\rho_x \in [0.1, 1]$ under design spaces
$$m\in \left[2,\frac{B/\underline{n} -c}{s}\right],~~~~n\in \left[6,\frac{B}{c+s\underline{m}}\right].$$ Figure \ref{fig:soMMD}
 (a) and (b) assume cluster-to-individual cost ratios of $k=10$ ($B=100,000$, $c=500$, $s=50$) and $k=20$ ($B=100,000$, $c=2,000$, $s=100$), respectively. A vertical dotted gray line depicts the maximin design; in the case of a cost ratio of $k=10$ the maximin design is $62$ clusters of size $22$, while in the $k=20$ case it is $18$ clusters of size $33$. Note that in each cost ratio case, the maximin design is at the intersection of the RE curves for ICC combinations ($\overline{\rho_{y|x}} =0.2$, $\underline{\rho_x}=0.1$) (dashed purple line) and ($\overline{\rho_{y|x}}=0.2$, $\overline{\rho_x}=1$) (dashed pink line). This makes intuitive sense as the LOD for the ($\overline{\rho_{y|x}}=0.2$, $\overline{\rho_x}=1$) scenario tends toward many small clusters so it reaches maximum RE early in the design space and then quickly becomes less relatively efficient as $m$ increases. On the other hand, in our example the LOD for the ($\overline{\rho_{y|x}} =0.2$, $\underline{\rho_x}=0.1$) scenario is the smallest number of large clusters possible within our constraints, so it is slow in reaching the maximum RE; its RE curve follows very closely to the ($\underline{\rho_{y|x}} =0.005$, $\underline{\rho_x}=0.1$) scenario (solid green line), which has the same LOD. Thus, it makes sense for the maximin design to be found at the intersection of scenarios that achieve their LOD most and least quickly, respectively.
 
The minimum RE for the maximin design in both cost-ratio scenarios (($m=22$, $n=62$) in the $k=10$ scenario, and ($m=33$, $n=18$) in the $k=20$ scenario) is approximately $0.68$. However, if the ICC combination(s) under which the maximin design is identified differs from the true ICC that generates the trial data, the RE of the maximin design may improve by as much as $32$\%. That is, if the true trial ICC combination is $(\rho_{y|x}=0.005, \rho_x=0.1)$ (Figure \ref{fig:soMMD}, solid green line), for example, instead of $(\rho_{y|x}=0.2, \rho_x=0.1)$ (dashed purple line) or $(\rho_{y|x}=0.2, \rho_x=1)$ (dashed pink line) then, given the maximin design is found at the intersection of the dashed purple and pink lines, the maximin design for either the k=10 or k=20 scenarios (($m=22$, $n=62$) and ($m=33$, $n=18$), respectively) can achieve a RE of approximately $0.90$ (and thus has a 32\% improvement over the minimum RE of 0.68). In addition, we confirm that the maximin design in the higher cost-ratio case favors a fewer number of large clusters compared to the lower cost-ratio case, reflecting the cost-effective strategy of expanding the cluster size to increase precision when recruiting an additional cluster becomes expensive and less practical. Of course, the corresponding overall or average treatment effect scenarios would result in much different maximin designs; for example, the average treatment effect maximin design for a cost ratio of $k=10$ would be $80$ clusters of size $15$, while it would be $23$ clusters of size $23$ in the $k=20$ case. In general, the ATE-oriented maximin design favors a greater number of smaller clusters compared to the HTE-oriented maximin design; this difference is because the HTE-oriented maximin design requires us to additionally consider the impact of covariate-ICC beyond the outcome-ICC.

\begin{figure}
    \centering
    \includegraphics[width=\textwidth]{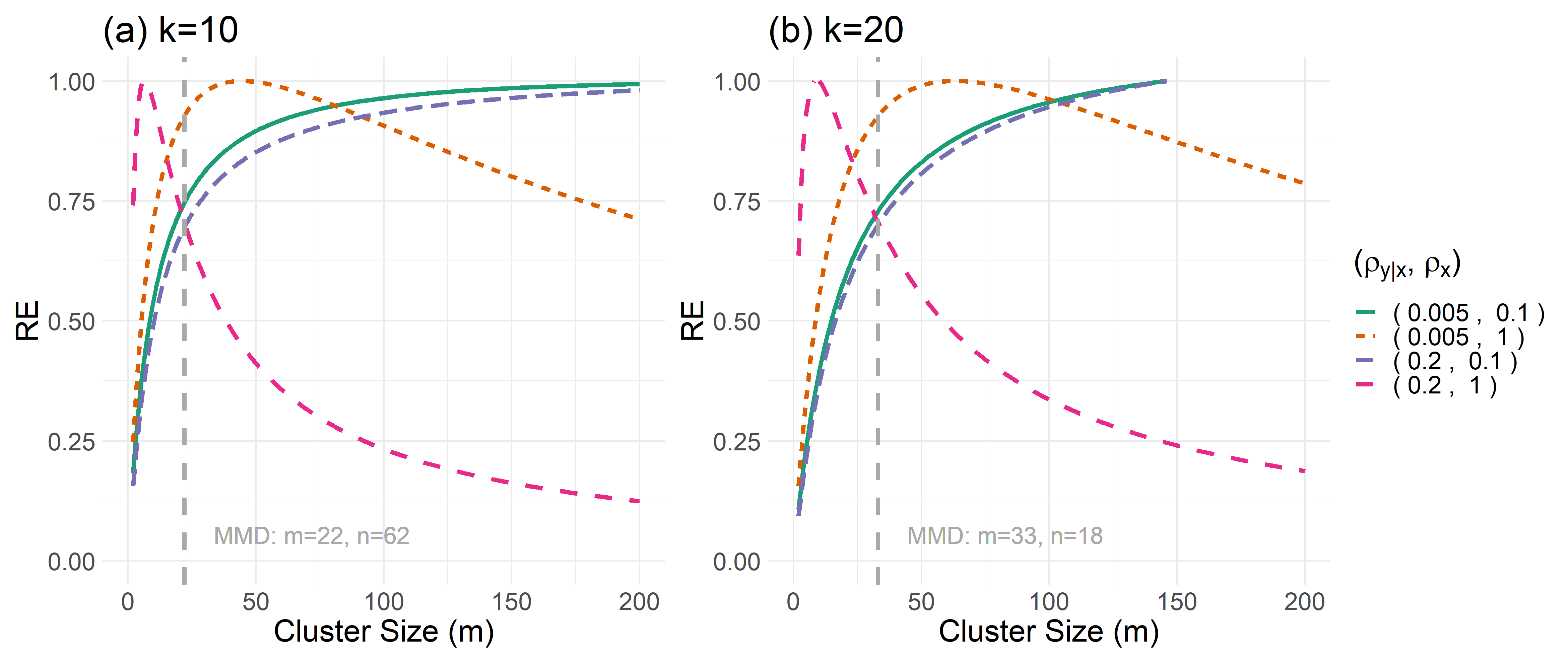}
    \caption{
     Plots of relative efficiencies (RE) of designs with cluster size $m$ versus their respective LODs for several ($\rho_{y|x}$, $\rho_x$) value combinations for a cluster-individual cost ratio of (a) $10$ and (b) $20$. The vertical dotted gray lines represent the maximin design (MMD) for assessing HTE in CRTs.
     }
\end{figure}

For completeness, we include three-dimensional RE plots in Appendix \ref{3dplots} for the $k=20$ case. The left panels of Figure C1 
illustrate the behavior of RE across the design space of $m$ and continuously across the parameter space of $\rho_{y|x}$ for fixed values of $\rho_x\in\{0.1, 0.5, 1\}$. The right panels of Figure C1 serve a similar purpose, but illustrate the behavior of RE continuously across the parameter space of $\rho_x$ for fixed values of $\rho_{y|x}\in\{0.005, 0.1, 0.2\}$. Dynamic versions of these plots can also be viewed via a freely-accessible R shiny web application at \url{https://mary-ryan.shinyapps.io/HTE-MMD-app/}.

\section{Optimal designs based on a compound optimality criterion}\label{s:mood}

In Section \ref{ss:solod}, the locally optimal and maximin designs are based on maximizing the power for detecting HTE, and are referred to as the \emph{single-objective designs}. In general, single-objective maximin designs are useful when we are only interested in powering a study with respect to a single analytic goal. The single-objective optimal design procedures developed for assessing the HTE, however, may or may not be optimal for assessing the average treatment effect as the respective estimators for these different effect measures have different variances relying on different sets of parameters. To balance the needs of these two objectives, in the following Section \ref{ss:molod} we construct a compound optimality criterion that allows us to find an optimal design taking into account both the average and heterogeneous treatment effect objectives assuming knowledge of the ICC parameters, and arrive at a multiple-objective locally optimal design. In Section \ref{ss:mommd},  we further extend this to the maximin design space to find a design that is optimal over a range of unknown ICC values. To encourage the exploration of a wider range of parameter spaces, we have also implemented the multiple-objective locally optimal and maximin designs in a freely-accessible R shiny web application at \url{https://mary-ryan.shinyapps.io/HTE-MMD-app/}.

\subsection{Locally optimal design}\label{ss:molod}
Let $\Theta_{\HTE}(\zeta)$ and $\Theta_{\ATE}(\zeta)$ denote the heterogeneous (minimize $\sigma^2_{\HTE}$) and average treatment effect (minimize $\sigma^2_{\ATE}$) objectives, respectively, under some design $\zeta$ in the design space. Similar to Moerbeek,\cite{moerbeek_optimal_2020} we create a compound function that takes both objectives into account:
\begin{equation}\label{eq:compoundObj}
    \Theta(\zeta|\lambda) = \lambda\Theta_{\ATE}(\zeta) + (1-\lambda)\Theta_{\HTE}(\zeta),
\end{equation}
where $\lambda\in[0,1]$ is a user-specified priority weight. As a linear combination of two objectives, this compound function includes two special cases when $\lambda$ takes the boundary values. That is, when $\lambda=0$, the objective function represents the efficiency objective for assessing HTE alone (and returns the methods in Section \ref{ss:solod}); when $\lambda=1$, the objective function coincides with the efficiency objective for assessing the average treatment effect alone (and returns some of the methods in Table \ref{tab:LODlit}, but replacing their marginal outcome-ICC with a conditional outcome-ICC). In other cases, assuming the average treatment effect objective will usually be the primary study priority, $\lambda$ can be specified such that the efficiency of $\Theta_{\HTE}(\zeta)$ is maximized while maintaining some minimal efficiency level for $\Theta_{\ATE}(\zeta)$, meaning $\lambda$ can be chosen as a value greater than $0.5$. In what follows, we will pursue locally optimal design assuming a fixed priority weight $\lambda$. 

Because the variance considered in each objective may be obtained on a different scale, we standardize each variance based on their respective LODs; for example, the LOD for assessing the HTE is derived in Proposition \ref{PROP:SOLOD}. Then our optimality criterion can be written as:
\begin{align}\label{eq:optCrit}
\begin{split}
    \min_m \Theta(\zeta|\lambda) &= \lambda \frac{\Theta_{\ATE}(\zeta)}{\Theta_{\ATE}(\zeta^*_{\ATE})} + (1-\lambda)\frac{\Theta_{\HTE}(\zeta)}{\Theta_{\HTE}(\zeta^*_{\HTE})}\\
    &= \frac{\lambda}{\Theta_{\ATE}(\zeta^*_{\ATE})}\sigma^2_{\ATE} + \frac{(1-\lambda)}{\Theta_{\HTE}(\zeta^*_{\HTE})}\sigma^2_{\HTE}\\
    &= \tilde{w}_{\ATE}\sigma^2_{\ATE} + \tilde{w}_{\HTE}\sigma^2_{\HTE},
\end{split}
\end{align}
where $\zeta^*_{O}$ represents the optimal design under objective $O\in\{\HTE,\ATE\}$, $\tilde{w}_O = \frac{\lambda}{\Theta_O(\zeta^*_O)}$ represents the weight $\sigma^2_O$ contributes to the criterion under objective $O\in\{\HTE,\ATE\}$, and ${\Theta_O(\zeta)}/{\Theta_O(\zeta^*_O)}$ can be interpreted as the inverse RE. Our goal is to minimize the compound objective to find the optimal design. This specification is similar to that used by Moerbeek.\cite{moerbeek_optimal_2020} In the current article, we instead propose to maximize the weighted combination of the REs to obtain the multiple-objective LOD, because RE (rather than inverse RE) is usually a more standard metric in deriving the optimal design. In our numerical explorations (results not shown), these two approaches frequently lead to similar optimal solutions, but RE criterion provides simpler and more regular solutions (solving quadratic functions rather than fourth-order polynomials). Specifically, we propose to solve for the optimal cluster size $m$ by maximizing the weighted combination of RE criterion:
\begin{align}\label{eq:optCritRE}
\begin{split}
    \max_m \Theta(\zeta|\lambda) &= \lambda \frac{\Theta_{\ATE}(\zeta^*_{\ATE})}{\Theta_{\ATE}(\zeta)} + (1-\lambda) \frac{\Theta_{\HTE}(\zeta^*_{\HTE})}{\Theta_{\HTE}(\zeta)}\\ 
    &= \lambda \Theta_{\ATE}(\zeta^*_{\ATE})\frac{1}{\sigma^2_{\ATE}} + (1-\lambda)\Theta_{\HTE}(\zeta^*_{\HTE})\frac{1}{\sigma^2_{\HTE}}\\
    &= \frac{w_{\ATE}}{\sigma^2_{\ATE}} + \frac{w_{\HTE}}{\sigma^2_{\HTE}}
\end{split}
\end{align}
where $\zeta^*_{O}$ is defined similar as in equation \eqref{eq:optCrit} and $w_O = \lambda\Theta_O(\zeta^*_O)$ represents the weight $1/\sigma^2_O$ contributes to the criterion under objective $O\in\{\HTE,\ATE\}$. The approach based on \eqref{eq:optCritRE} has the benefit of greater interpretability and a more elegant closed-form solution for the multiple-objective LOD, which we outline in Proposition \ref{PROP:MOLOD}. 

\begin{proposition}\label{PROP:MOLOD}
\textit{Let the compound optimality criterion for the average and heterogeneous treatment effect objectives be defined as in \eqref{eq:optCritRE} and given by:}
\begin{equation*}
    \Theta(\zeta|\lambda) = \lambda \frac{\Theta_{\ATE}(\zeta^*_{\ATE})}{\Theta_{\ATE}(\zeta)} + (1-\lambda) \frac{\Theta_{\HTE}(\zeta^*_{\HTE})}{\Theta_{\HTE}(\zeta)} = \frac{w_{\ATE}}{\sigma^2_{\ATE}} + \frac{w_{\HTE}}{\sigma^2_{\HTE}}.
\end{equation*}
\textit{Then, given a budget constraint \eqref{eq:budget}, well-defined outcome- and covariate-ICCs, and a priority weight $\lambda$, the locally optimal design for a cluster randomized design that maximizes this compound criterion is given by:}
\begin{equation}\label{eq:moLOD}
    m_{\opt} = \frac{-w_{\HTE}ka_2 - \sqrt{w^2_{\HTE}k^2a^2_2 - 4\left\{w_{\HTE}(ka_1 - b_1) - w_{\ATE}\rho_{y|x}\right\}\left\{w_{ATE}k(1-\rho_{y|x}) + w_{\HTE}ka_3\right\}}}{2\left\{w_{\HTE}(ka_1 - b_1) - w_{\ATE}\rho_{y|x}\right\}},
\end{equation}
\textit{under the condition that}
\begin{equation}\label{eq:moLODCond}
    w_{\ATE} > w_{\HTE} \left\{(k+1)\rho_{y|x} - \rho_x(k\rho_{y|x}+1)\right\}\text{ and } m_{\opt} \le \frac{B/\underline{n} - c}{s},
\end{equation}
where $a_1=\rho^2_{y|x}(1-\rho_x)$, $a_2=2\rho_{y|x}(1-\rho_{y|x})(1-\rho_x)$, $a_3=(1-2\rho_{y|x}+\rho_x\rho_{y|x})(1-\rho_{y|x})$, $b_1=\rho_{y|x}(\rho_x-\rho_{y|x})$. \textit{If condition \eqref{eq:moLODCond} is not satisfied, then we set}
\begin{equation*}
    m_{\opt} = \frac{B/\underline{n} - c}{s}.
\end{equation*}
\textit{In either case, the optimal number of clusters is given by}
\begin{equation*}
    n_{\opt} = \frac{B}{c+sm_{\opt}}
\end{equation*}

\begin{proof}
See Appendix \ref{moLODProof}.
\end{proof}
\end{proposition}

It is worth noting that in the case where covariates are collected at the cluster level ($\rho_x=1$), the multiple-objective LOD given by Proposition \ref{PROP:MOLOD} coincides with the single-objective LOD for assessing treatment effect heterogeneity given in Proposition \ref{PROP:SOLOD}, as well as with the single-objective LOD for the average treatment effect. In addition, our numerical explorations suggest that the condition \eqref{eq:moLODCond} is satisfied under a wide range of the parameter space, and is only likely a critical condition when $\lambda<0.25$, in which case the priority weight largely favors the objective to assess treatment effect heterogeneity.

To explore the pattern of the multiple-objective LOD, Table \ref{tab:moLOD} presents several examples obtained by calculating \eqref{eq:moLOD}. We assume $k=10$ and a minimum of $\underline{n}=6$ clusters; we additionally assume we are powering both the average treatment effect and HTE for equal standardized effect sizes of $\beta_2/\sigma_{y|x}=\beta_4\sigma_x/\sigma_{y|x}=0.2$ ($\sigma^2_{y|x} = \sigma^2_x = 1$). Multiple-objective LODs were then found assuming priority weights of $\lambda\in\{0.4,0.6,0.85\}$. As $\lambda\rightarrow 1$, the multiple-objective LODs allocate more power to the average treatment effect objective than the HTE objective compared to the LOD for the same ICC values under a smaller $\lambda$. We also observe that the LOD changes less with shifts in $\rho_x$ and fixed $\rho_{y|x}$ when $\lambda=0.85$ and tends toward a larger number of smaller clusters than under $\lambda=0.6$ or $0.4$; this is because more weight is being given to the average treatment effect objective, which is less sensitive to changes in $\rho_x$ as the covariate-ICC does not factor into $\sigma^2_{\ATE}$. More substantial changes in the LOD are seen with shifts of $\rho_{y|x}$ when $\lambda=0.85$. As $\lambda \rightarrow 0$, we see there are more gradual changes in the LOD for shifts of $\rho_{y|x}$ and $\rho_x$, moving more evenly from a smaller number of large clusters to many small clusters as both ICCs increase; this is because priority is shifted to the HTE objective, whose variance depends on both ICC parameters. We also see that this shift toward the HTE objective results in the LODs favoring fewer, larger clusters than for greater values of $\lambda$ at the same $(\rho_{y|x}, \rho_x)$. Finally, for a fixed $\rho_{y|x}$ and $\lambda$, the power for the average treatment effect remains fairly stable as $\rho_x$ increases while the power for assessing HTE can vary to a greater degree, especially if the value of the outcome-ICC, $\rho_{y|x}$, is large.

\begin{table}
\caption{\label{tab:moLOD}Locally optimal cluster size ($m$),  number of clusters ($n$), and power to detect standardized average (ATE) and heterogeneous treatment effect (HTE) sizes of $0.2$ for known outcome-ICC ($\rho_{y|x}$) and covariate-ICC ($\rho_x$) values assuming a total budget $B=100,000$, cluster-associated costs $c=500$, individual-associated costs $s=50$ (cost ratio $k=10$), and $\sigma^2_{y|x} = \sigma^2_x = 1$.}
\centering
\begin{tabular}{ll | llll | llll | llll}
\toprule
&& \multicolumn{4}{c}{$\lambda = 0.4$} & \multicolumn{4}{c}{$\lambda = 0.6$} & \multicolumn{4}{c}{$\lambda=0.85$}\\
\midrule
 && && \multicolumn{2}{c|}{\emph{Power}} &&& \multicolumn{2}{c|}{\emph{Power}} &&& \multicolumn{2}{c}{\emph{Power}}\\ 
$\rho_{y|x}$ & $\rho_x$ & $m$ & $n$ & ATE & HTE & $m$ & $n$ & ATE & HTE & $m$ & $n$ & ATE & HTE\\
\midrule
0.005 & 0.1 & 72 & 24 & 0.947 & 0.984 & 58 & 29 & 0.952 & 0.982 & 48 & 34 & 0.954 & 0.979\\
& 0.2 & 68 & 25 & 0.947 & 0.980 & 56 & 30 & 0.953 & 0.980 & 48 & 34 & 0.954 & 0.977\\
& 0.5 & 57 & 29 & 0.950 & 0.970 & 52 & 32 & 0.955 & 0.972 & 46 & 35 & 0.953 & 0.969\\
& 0.75 & 50 & 33 & 0.954 & 0.963 & 48 & 34 & 0.954 & 0.963 & 45 & 36 & 0.955 & 0.963\\
& 1 & 44 & 37 & 0.956 & 0.955 & 44 & 37 & 0.956 & 0.955 & 44 & 37 & 0.956 & 0.955\\
\midrule
0.05 & 0.1 & 26 & 55 & 0.735 & 0.961 & 18 & 71 & 0.769 & 0.942 & 15 & 80 & 0.778 & 0.929\\
& 0.2 & 25 & 57 & 0.743 & 0.950 & 18 & 71 & 0.769 & 0.931 & 14 & 83 & 0.777 & 0.910\\
& 0.5 & 21 & 64 & 0.758 & 0.893 & 17 & 74 & 0.774 & 0.883 & 14 & 83 & 0.777 & 0.867\\
& 0.75 & 17 & 74 & 0.774 & 0.829 & 15 & 80 & 0.778 & 0.824 & 14 & 83 & 0.777 & 0.819\\
& 1 & 13 & 86 & 0.774 & 0.753 & 13 & 86 & 0.774 & 0.753 & 13 & 86 & 0.774 & 0.753\\
\midrule
0.1 & 0.1 & 19 & 68 & 0.62 & 0.949 & 12 & 90 & 0.667 & 0.908 & 10 & 100 & 0.677 & 0.885\\
& 0.2 & 19 & 68 & 0.62 & 0.934 & 12 & 90 & 0.667 & 0.891 & 10 & 100 & 0.677 & 0.868\\
& 0.5 & 16 & 76 & 0.642 & 0.949 & 12 & 90 & 0.667 & 0.821 & 10 & 100 & 0.677 & 0.802\\
& 0.75 & 12 & 90 & 0.667 & 0.736 & 11 & 95 & 0.673 & 0.734 & 10 & 100 & 0.677 & 0.727\\
& 1 & 9 & 105 & 0.676 & 0.630 & 9 & 105 & 0.676 & 0.630 & 9 & 105 & 0.676 & 0.630\\
\midrule
0.2 & 0.1 & 13 & 86 & 0.527 & 0.937 & 8 & 111 & 0.576 & 0.870 & 6 & 125 & 0.581 & 0.806\\
& 0.2 & 13 & 86 & 0.527 & 0.917 & 8 & 111 & 0.576 & 0.846 & 6 & 125 & 0.581 & 0.782\\
& 0.5 & 11 & 95 & 0.550 & 0.799 & 8 & 111 & 0.576 & 0.750 & 6 & 125 & 0.581 & 0.694\\
& 0.75 & 9 & 105 & 0.568 & 0.646 & 7 & 117 & 0.578 & 0.619 & 6 & 125 & 0.581 & 0.602\\
& 1 & 6 & 125 & 0.581 & 0.491 & 6 & 125 & 0.581 & 0.491 & 6 & 125 & 0.581 & 0.491\\
\bottomrule
\end{tabular}
\end{table}

\subsection{Maximin design}\label{ss:mommd}

In Table \ref{tab:moLOD}, we observe that the multiple-objective LOD varies for different combinations of outcome-ICC and covariate-ICC. Thus, we can use the compound optimality criterion \eqref{eq:optCritRE} within a maximin design framework to find a design robust to ICC misspecification that appropriately powers both the average and heterogeneous treatment effect objectives within the given budget constraints. This multiple-objective maximin design procedure is summarized in Algorithm \ref{algo:moMMD}.

\begin{algorithm}
\caption{Multiple-objective maximin design procedure based on the compound optimality criterion}\label{algo:moMMD}
\begin{algorithmic}[1]
    \State Choose priority weight $\lambda$;
    \State Define the parameter ($\rho_{y|x}$, $\rho_x$) and design $\left(m, n(m)\right)$ spaces;
    \State For each ($\rho_{y|x}$, $\rho_x$) parameter value combination, compute the LOD for each objective based on Proposition \ref{PROP:SOLOD} and methods (for assessing the average treatment effect) in Table \ref{tab:LODlit}. Then compute the compound optimality criterion $\Theta(\zeta|\lambda)$ for each $\left(m, n(m)\right)$ design value combination compared with the LODs at the parameter value pair according to \eqref{eq:optCritRE};
    \State For each $\left(m, n(m)\right)$ design value combination, choose the ($\rho_{y|x}$, $\rho_x$) parameter value combination that has the smallest criterion value;
    \State Among the smallest criterion values, choose the $\left(m, n(m)\right)$ design value combination that has the largest criterion value.
\end{algorithmic}
\end{algorithm}

\begin{figure}
\centering
\includegraphics[width=0.95\textwidth]{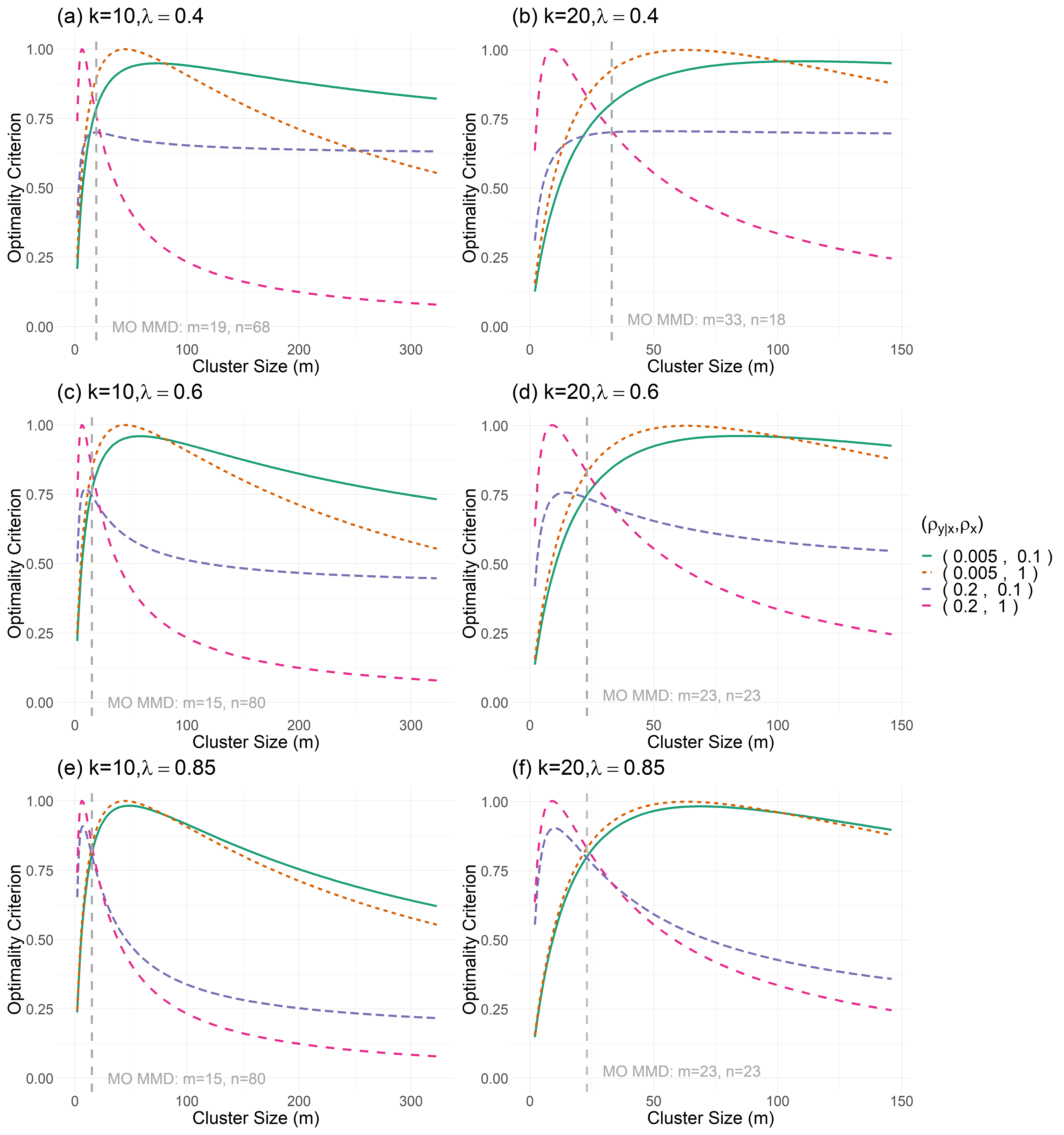}
\raggedleft\caption{\label{fig:moMMD}Plots of weighted REs of designs with cluster size $m$ versus their respective LODs for several ($\rho_{y|x}$, $\rho_x$) value combinations for a cluster-individual cost ratio of $10$  (Panels: a,c,e) and  $20$ (Panels: b,d,f) and priority weights  $\lambda=0.4$ (Panels: a,b),  $\lambda=0.6$ (Panels: c,d), and  $\lambda=0.85$ (Panels: e,f). The vertical dotted gray lines represent the multiple-objective maximin design (MO MMD).}
\end{figure}

Figure \ref{fig:moMMD} illustrates examples of multiple-objective maximin designs for cost ratios of $k=10$ (left panels) and $k=20$ (right panels) as well as priority weights $\lambda=0.4$ (top row), $\lambda=0.6$ (middle row), and $\lambda=0.85$ (bottom row). We explored the following parameter and design spaces:
\begin{equation*}
\rho_{y|x}\in [0.005, 0.2],~~~~\rho_x\in [0.1, 1],~~~~m\in \left[2,\frac{B/\underline{n}-c}{s}\right],~~~~n\in\left[6, \frac{B}{c+s\underline{m}}\right].
\end{equation*}
Unlike in the single-objective maximin design case where the maximin design was often the intersection of RE curves from two ICC value combinations, in the compound objective case there are often three different ICC value combinations that achieve local optimality criterion minimums on portions of the range of $m$ and the maximin design generally falls somewhere on 
the range of the ``middle'' local minimum; this is seen most clearly when $k=20$ and $\lambda=0.4$ and $0.6$ (Figure \ref{fig:moMMD} panels (b) and (d)) where the ``middle'' local minimum refers to the dashed purple line for ICC scenario $(0.2, 0.1)$. In our particular example, regardless of cost ratio, we see that the maximin design is found at the intersection of $(\overline{\rho_{y|x}}, \underline{\rho_x})$ (dashed purple line) and $(\overline{\rho_{y|x}}, \overline{\rho_x})$ (dashed pink line) when priority is placed on the HTE objective ($\lambda < 0.5$; Figure \ref{fig:moMMD} (a)-(b)); this is the same scenario intersection that the single HTE-objective maximin design was found at in Figure \ref{fig:soMMD} (Figure \ref{fig:soMMD} can be thought of as an extreme $\lambda=0$ case of the multiple-objective maximin design). On the other hand, when priority is placed on the average treatment effect objective ($\lambda > 0.5$; Figure \ref{fig:moMMD} (c)-(f)) the maximin design is found at the intersection of $(\overline{\rho_{y|x}}, \underline{\rho_x})$ (dashed purple line) and $(\underline{\rho_{y|x}}, \underline{\rho_x})$ (solid green line). We also see that as $\lambda \rightarrow 1$ the curves for the $(\underline{\rho_{y|x}}, \underline{\rho_x})$ and $(\underline{\rho_{y|x}}, \overline{\rho_x})$ (solid green and dotted orange) scenarios converge, as do the curves for $(\overline{\rho_{y|x}}, \underline{\rho_x})$ and $(\overline{\rho_{y|x}}, \overline{\rho_x})$ (dashed purple and dashed pink) scenarios; this is because $\sigma^2_{\ATE}$ only depends on $\rho_{y|x}$ and therefore the pure ATE-oriented maximin design is found at the intersection of boundary values of the outcome-ICC, regardless of the value of the covariate-ICC. In addition, while the multiple-objective maximin design can vary across cost ratios, it does not vary excessively across values of $\lambda$, especially when comparing across different values of $\lambda$ giving ``majority weight'' to the same objective. 

As in the single-objective case, we include three-dimensional optimality criterion plots in Appendix \ref{3dplots} for the $k=20$ and $\lambda=0.6$ case. The left panels of Figure C2 
illustrate the behavior of optimality criterion across the design space of $m$ and continuously across the parameter space of $\rho_{y|x}$ for fixed values of $\rho_x\in\{0.1, 0.5, 1\}$. The right panels of Figure C2 serve a similar purpose, but illustrating the behavior of optimality criterion continuously across the parameter space of $\rho_x$ for fixed values of $\rho_{y|x}\in\{0.005, 0.1, 0.2\}$. Dynamic versions of these plots can also be viewed via a freely-accessible R shiny web application at \url{https://mary-ryan.shinyapps.io/HTE-MMD-app/}.

\section{Power Considerations for Maximin Designs}\label{s:power}

The maximin design procedures proposed in this article allow investigators to identify an optimal study sample size in the face of uncertainty regarding outcome- and covariate-ICC values at the design stage. Once an optimal design is found using the proposed maximin design procedures, though, a question still remains as to how one might conduct power calculations for the next stage in study planning, including decisions around ICC values to use. While our main focus in this work is on the identification of optimal designs based on relative efficiency, we provide some perspectives on power calculation for completeness.

\begin{figure}
\centering
\includegraphics[width=1\textwidth]{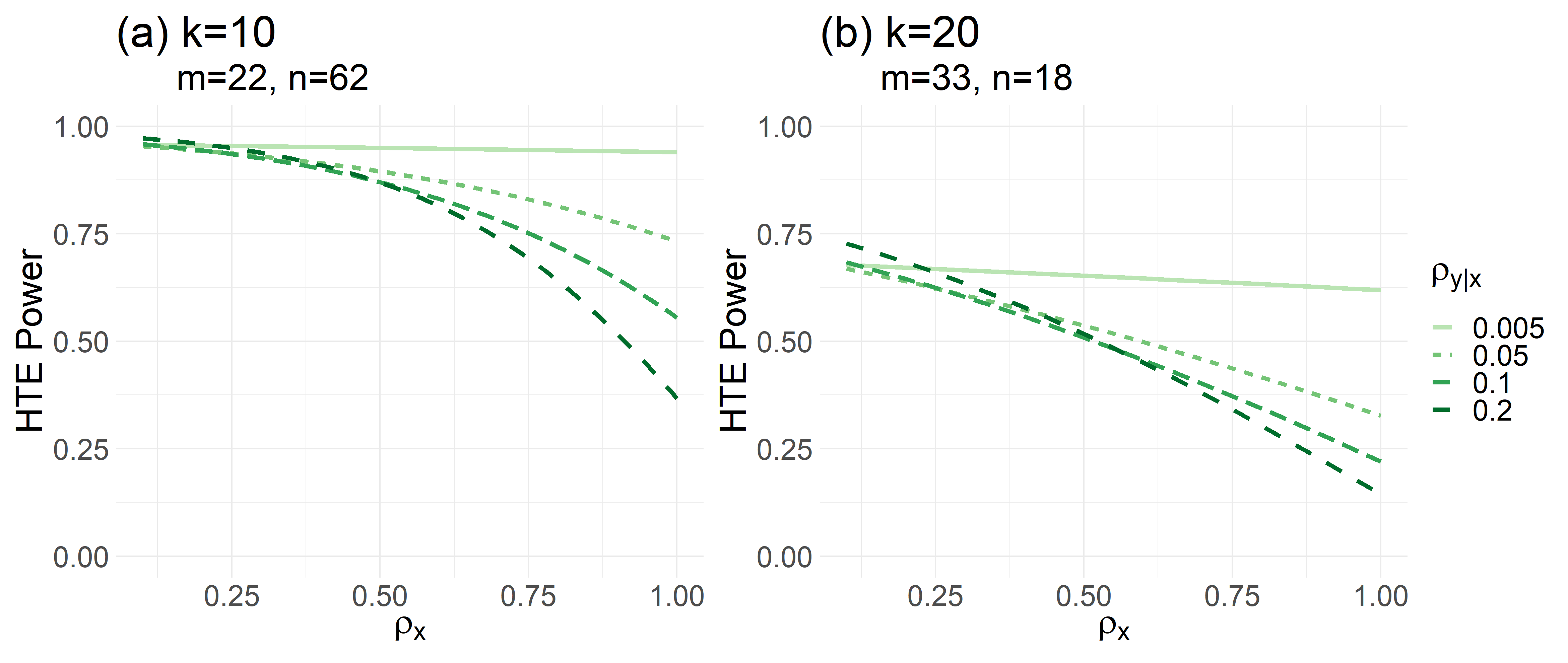}
\raggedleft\caption{\label{fig:soMMD-power}Power curves for a standardized HTE effect size of $0.2$ across $\rho_x$ for four $\rho_{y|x}$ values  for a cluster-individual cost ratio of $10$  (a) and  $20$ (b), evaluated at their respective maximin designs.}
\end{figure}

We begin by assuming that investigators are interested in exploring power over the same ICC parameter space as they previously used for identification of the maximin design. Next, as uncertainty around outcome- and covariate-ICC values still remains for study investigators, it may benefit investigators to calculate power under a grid search of the ICC parameter space as sensitivity analyses. As an example, power curves for the single HTE objective with standardized effect size of $\beta_4\sigma_x/\sigma_{y|x}=0.2$ ($\sigma^2_{y|x} = \sigma^2_x = 1$) and at cost ratios $k=10$ (a) and $k=20$ (b), evaluated at their respective maximin designs identified in Section \ref{ss:sommd}, are shown in Figure \ref{fig:soMMD-power}. Similar power curves assessing power for the heterogeneous and average treatment effects evaluated at maximin designs identified in Section \ref{ss:mommd} at $\lambda=0.6$ are shown in Figure C3 in Appendix \ref{3dplots}.

Overall we observe that, at a particular maximin design and for a fixed $\rho_{y|x}$, power of the HTE test decreases as $\rho_x$ increases. When $\rho_x$ is very small, higher power will be achieved under larger $\rho_{y|x}$; when $\rho_x$ is large, higher power is obtained under small $\rho_{y|x}$. Power differences between $\rho_{y|x}$ at small $\rho_x$ will be more stark at large cost ratios $k$ (panel b). We observe that, in general, the highest HTE power is attained at $(\overline{\rho_{y|x}}, \underline{\rho_x})$ while the lowest power is attained at $(\overline{\rho_{y|x}}, \overline{\rho_x})$; this reflects previous results regarding the parabolic relationship between $\rho_{y|x}$ and the variance of the HTE estimator for fixed, non-optimal designs.\cite{yang_sample_2020} We note that the endpoints of the lightest and darkest lines in Figure \ref{fig:soMMD-power} represent the boundary ICC combinations that were assessed at the maximin design stage. Thus, we can then establish a lower bound for the HTE power of our maximin design at $(\overline{\rho_{y|x}}, \overline{\rho_x})$ and an upper bound at $(\overline{\rho_{y|x}}, \underline{\rho_x})$. In the case of our examples in Figure \ref{fig:soMMD-power}, we would conclude that our maximin design would have power to detect a standardized $\beta_4\sigma_x/\sigma_{y|x}=0.2$ as low as $36.7$\% and as high as $97.2$\% when $k=10$, and as low as $14.4$\% and as high as $72.8$\% when $k=20$. These power bounds, along with accompanying power curves such as those shown in Figure \ref{fig:soMMD-power}, can be used by investigators to assess whether satisfactory power is achieved under the allocated budget across a range of ICC values. We make an additional note that, in the case of assessing power for the average treatment effect after identifying the optimal sample size via the multiple-objective maximin design (as we do in Figure C3 in Appendix \ref{3dplots}), one needs only $\rho_{y|x}$ to create upper and lower power bounds. This is because $\sigma^2_{\ATE}$ does not involve $\rho_x$. In addition, the power of the average treatment effect test generally decreases with increasing values of $\rho_{y|x}$, and hence the lower and upper bound of $\rho_{y|x}$ often corresponds to the upper and lower bound of power of the average treatment effect test.

In our evaluation of optimal designs and study power, we have primarily focused on studying the impact of changes in each single design parameter while holding the other design parameters constant. Our goal has been to assess, in the study planning stage, how changes in the resulting optimal design and study power may be sensitive to input values for each individual design parameter (and therefore understand the anticipated trend), rather than to indicate that the design parameters are variationally independent in practice. In practice, it may not always be the case that one ICC parameter will stay fixed if the other is increased or decreased due to the implicit relationship between $\rho_x$ and $\rho_{y|x}$. For example, moving to a more homogeneous population of covariate $X$ (increasing $\rho_x$) may not affect the marginal homogeneity of the outcome ($\rho_y$)  but may decrease the homogeneity of the outcome conditional on the covariate ($\rho_{y|x}$) if the covariate is highly correlated with the outcome (due to explained variation).

Finally, we acknowledge that the total budget provided is often a key consideration to ensure a practical design where the testing objectives are properly powered in the worst case scenario. If either the single- or multiple-objective maximin design provides insufficient power for the desired effect size(s) under the worst case scenario regarding the ICC assumptions, the total budget must be increased to provide a larger statistical power under the worst case scenario. In addition, power will also depend on the assumptions on the range for ICC values. If pilot or routinely-collected data are available to help elicit narrower ranges of the covariate- or outcome-ICC values, that information should inform the power calculation for the obtained maximin design, and can often improve the power under the worst case scenario compared to using unnecessarily wide ICC ranges. There are recent efforts that report outcome-ICCs for cluster randomized trials,\cite{korevaar_intra-cluster_2021} and we encourage similar efforts to report covariate-ICCs for planning CRTs to detect treatment effect heterogeneity. 

\section{Application to the Kerala Diabetes Prevention Program (K-DPP) Study}\label{s:dataApp}

We illustrate our single- and multiple-objective optimal design procedures using data from the Kerala Diabetes Prevention Program (K-DPP) study,\cite{thankappan_peer-support_2018} a cluster-randomized controlled trial of a peer-support lifestyle intervention to reduce progression to diabetes in a community setting in India; we use data that are publicly available from the figshare database: \url{https://figshare.com/articles/dataset/K-DPP_datasets/5661610}. In the actual study, participants at high-risk for diabetes were recruited from $60$ polling areas (clusters) in a subdistrict of Kerala state, and polling areas were randomized in a $1$:$1$ ratio to receive usual care (education booklet on general lifestyle advice) or a $12$-month peer-support lifestyle intervention consisting of $15$ group sessions primarily led by trained lay peer leaders and held in local neighborhood facilities. The intervention was specifically designed to reduce cost and resource burden so as to be more readily employed in low- and middle-income countries where diabetes incidence is on the rise. The primary outcome was incidence of diabetes at $24$ months, but secondary outcomes included change in Indian Diabetes Risk Score (IDRS). Post-hoc HTE subgroup analyses were also conducted based on baseline glucose tolerance group, including impaired fasting glucose (IFG) as defined by the World Health Organization.

In the context of the K-DPP study, suppose study investigators are interested in conducting a CRT to evaluate the benefit of the peer-support lifestyle intervention among the population at high-risk for developing diabetes as measured by change in IDRS, as well as to see if such benefit is differential by baseline body mass index (BMI) and IFG status. As the randomization ratio is $1$:$1$, the variance of the treatment variable is given by $\sigma^2_w=0.25$. The original study reported cluster-level costs in the intervention arm to be approximately \$$241.20$ per cluster (personnel, travel, food, logistics, and communication costs for training and group sessions) and individual-level costs to be approximately \$$7.98$ per participant, (resource materials and administrative costs), with the intervention arm costing a total of \$$11,225$ in 2013 USD; this results in a cluster-to-individual cost ratio of $k\approx 30$. Assuming cluster- and individual-level costs in the control arm were considerably less, we will assume a total budget of \$$20,000$ and a global cost ratio of $k=20$ ($c=\$100$, $s=\$5$) for our study planning; we will discuss extension to heterogeneous cost ratios in Section \ref{s:discuss}. The original study also reported that peer groups had approximately $10$ to $23$ participants each; therefore we will restrict our search for optimal cluster size $m$ to be in the space $[8, 40]$ and total clusters $n$ to be in $[66,143]$. Based on publicly available K-DPP data, we estimate the marginal standard deviation of change in IDRS to be $\sigma_y=10.270$. The original study did not report observed outcome-ICC but we estimate it using a linear mixed model procedure to be $0.028$ conditional on BMI and 0.032 conditional on IFG; for the maximin design procedures, we will define the parameter space for $\rho_{y|x}$ to be $[0.005, 0.1]$. The K-DPP study observed an average treatment effect size of $\beta_2=-1.50$ IDRS points. From baseline K-DPP data, the mean BMI is $24.888$ ($\sigma_x=4.031$) and approximately $22.5$\% of participants have IFG ($\sigma_x=0.417$). We estimate the covariate-ICC for BMI to be $0.055$ ($95$\% CI: $[0.022, 0.106]$)\cite{mcgraw_forming_1996} and for IFG to be $0.012$ ($95$\% CI: $[0,0.093]$),\cite{zou_confidence_2004} where the ICC for IFG was estimated using the ANOVA method.\citep{ridout_estimating_1999} We will define the parameter space for both $\rho_x$ to be $[0.1, 0.75]$, as we likely would not have had a precise parameter range at the design stage. Suppose we are interested in an effect size of the treatment-by-BMI interaction of $\beta_4=0.25\times\beta_2=-0.375$ points, and an effect size for the IFG interaction that is the same as the average treatment effect. In practice, the HTE effect size associated with IFG may be smaller, but we select a relatively large effect size to offset the relatively small IFG covariate variance (as a binary effect modifier). Because we will perform design calculations based on both the ATE and HTE objectives, we assume the BMI and IFG are mean centered.

If we were solely interested in powering for the one of the HTEs, the LOD for optimally testing the HTE with respect to either BMI or IFG is obtained as one with $66$ clusters of size $40$ (total sample size $N=2,640$), as we are assuming a minimum of $66$ clusters. Using Algorithm \ref{algo:soMMD}, we find that the maximin design agrees with the LOD: the optimal design would be one with $66$ clusters of size $40$, found at ($\overline{\rho_{y|x}}=0.1, \underline{\rho_x}=0.1$), giving us $96.5$\% power to detect the HTE with respect to BMI at that scenario and $69.7$\% power at $(\overline{\rho_{y|x}}=0.1, \overline{\rho_x}=0.75)$, the scenario with the worst power. On the other hand, we get $34.6$\% power to detect the HTE with respect to IFG under the ($\overline{\rho_{y|x}}=0.1, \underline{\rho_x}=0.1$) scenario and 17.4\% power at $(\overline{\rho_{y|x}}=0.1, \overline{\rho_x}=0.75)$. At the observed ICC values, ($\rho_{y|x}=0.028, \rho_{\BMI}=0.055$) and ($\rho_{y|x}=0.032, \rho_{\IFG}=0.012$), we have $96.5$\% power to detect the HTE with respect to BMI and $34.9$\% power to detect the HTE with respect to IFG. These apparent differences in power between the BMI and IFG effects are mainly due to differences in the magnitude of the standard deviations of the effect modifier.

Table \ref{tab:moLOD-KDPP} further shows the multiple-objective LODs obtained when the true ICCs are known, ($\rho_{y|x}=0.028$, $\rho_{\BMI}=0.055$) and ($\rho_{y|x}=0.032$, $\rho_{\IFG}=0.012$), with priority weights varying between $0.5$ and $0.95$ (recall that as the priority weight goes to $1$, the objective criterion favors the objective for studying the average treatment effect). As the covariate-ICCs for BMI and IFG are not vastly different, their respective multiple-objective LODs are also found to be relatively similar.

\begin{table}
\caption{\label{tab:moLOD-KDPP}LOD, value of the optimality criterion, and power to detect an average treatment effect size of $-1.5$ and either a BMI HTE effect size of $-0.375$ or an IFG effect size of $-1.5$ at various $\lambda$ values assuming a total budget $B=20,000$, cluster-associated costs $c=100$, and individual-associated costs $s=5$ (cost ratio $k=20$).}
\centering
\begin{tabular}{l | cllll | cllll}
\toprule
 &\multicolumn{5}{c|}{\emph{BMI}} & \multicolumn{5}{c}{\emph{IFG}}\\ 
$\lambda$ & Optimality Criterion & $m$ & $n$ & ATE & HTE & Optimality Criterion & $m$ & $n$ & ATE & HTE\\
\midrule
0.5 & 0.827 & 42 & 64 & 0.747 & 0.967 & 0.814 & 40 & 66 & 0.723 & 0.349\\
0.55 & 0.839 & 38 & 68 & 0.753 & 0.962 & 0.827 & 36 & 71 & 0.734 & 0.340\\
0.6 & 0.854 & 36 & 71 & 0.760 & 0.960 & 0.843 & 34 & 74 & 0.741 & 0.335\\
0.65 & 0.869 & 33 & 75 & 0.764 & 0.955 & 0.860 & 32 & 76 & 0.740 & 0.326\\
0.7 & 0.886 & 32 & 76 & 0.763 & 0.952 & 0.878 & 30 & 80 & 0.748 & 0.322\\
0.75 & 0.903 & 30 & 80 & 0.771 & 0.949 & 0.900 & 29 & 81 & 0.746 & 0.316\\
0.8 & 0.922 & 29 & 81 & 0.768 & 0.945 & 0.916 & 28 & 83 & 0.748 & 0.314\\
0.85 & 0.941 & 28 & 83 & 0.770 & 0.943 & 0.937 & 27 & 85 & 0.750 & 0.310\\
0.9 & 0.960 & 27 & 85 & 0.772 & 0.941 & 0.957 & 26 & 86 & 0.747 & 0.304\\
0.95 & 0.980 & 26 & 86 & 0.768 & 0.935 &  0.979 & 25 & 88 & 0.747 & 0.299\\
\bottomrule
\end{tabular}
\end{table}

Designs obtained using multiple-objective maximin design Algorithm \ref{algo:moMMD} and varying the priority weight $\lambda$ between $0.5$ and $0.95$ are summarized in Table \ref{tab:moMMD-KDPP}. If the average and heterogeneous treatment effect objectives are given equal priority, the optimal design is one with $86$ clusters of size $26$ (total sample size $N=2,236$) found at the intersection of ($\rho_{y|x}=0.005$, $\rho_x=0.1$) and ($\rho_{y|x}=0.1$, $\rho_x=0.1$). If the average treatment effect objective is given a priority of at least $\lambda=0.6$, the optimal design becomes the one with $85$ clusters of size $27$ (total sample size $N=2,295$) also found at the intersection of ($\rho_{y|x}=0.005$, $\rho_x=0.1$) and ($\rho_{y|x}=0.1$, $\rho_x=0.1$). This gives us between $49.7$\% and $91.2$\% power to detect the average treatment effect, between $68.7$\% and $94.2$\% power to detect the HTE with respect to BMI, and between $17.1$\% and $30.7$\% power to detect the HTE with respect to IFG. At the observed ICC values, ($\rho_{y|x}=0.028$, $\rho_{\BMI}=0.055$) and ($\rho_{y|x}=0.032$, $\rho_{\IFG}=0.012$), we have $77.2$\% power to detect the average treatment effect, $94.1$\% power to detect the HTE with respect to BMI, and $31.0$\% power to detect the HTE with respect to IFG.

\begin{table}
\caption{\label{tab:moMMD-KDPP}Maximin design, value of the optimality criterion, and upper and lower power bounds to detect an average treatment effect (ATE) size of $-1.5$ and either a BMI HTE effect size of $-0.375$ or an IFG effect size of $-1.5$ at various $\lambda$ values assuming a total budget $B=20,000$, cluster-associated costs $c=100$, and individual-associated costs $s=5$ (cost ratio $k=20$).}
\centering
\begin{tabular}{l | cllccc}
\toprule
 &&&& \multicolumn{3}{c}{Power Bounds}\\ \cmidrule{5-7}
$\lambda$ & Optimality Criterion & $m$ & $n$ & ATE & HTE (BMI)  & HTE (IFG)\\
\midrule
0.5 & 0.742 & 26 & 86 & (0.497 - 0.906) & (0.681 - 0.937) & (0.169 - 0.301)\\
0.55 & 0.757 & 26 & 86 & (0.497 - 0.906) & (0.681 - 0.937) & (0.169 - 0.301)\\
0.6 & 0.772 & 27 & 85 & (0.497 - 0.912) & (0.687 - 0.943) & (0.171 - 0.308)\\
0.65 & 0.787 & 27 & 85 & (0.497 - 0.912) & (0.687 - 0.943) & (0.171 - 0.308)\\
0.7 & 0.801 & 27 & 85 & (0.497 - 0.912) & (0.687 - 0.943) & (0.171 - 0.308)\\
0.75 & 0.816 & 27 & 85 & (0.497 - 0.912) & (0.687 - 0.943) & (0.171 - 0.308)\\
0.8 & 0.828 & 27 & 85 & (0.497 - 0.912) & (0.687 - 0.943) & (0.171 - 0.308)\\
0.85 & 0.841 & 27 & 85 & (0.497 - 0.912) & (0.687 - 0.943) & (0.171 - 0.308)\\
0.9 & 0.853 & 27 & 85 & (0.497 - 0.912) & (0.687 - 0.943) & (0.171 - 0.308)\\
0.95 & 0.867 & 28 & 83 & (0.491 - 0.914) & (0.687 - 0.945) & (0.171 - 0.311)\\
\bottomrule
\end{tabular}
\end{table}

\section{Discussion}\label{s:discuss}

Interest in assessing differential treatment effects among subpopulations, in addition to assessing overall treatment effect, is increasing in the setting of CRTs. Understanding treatment effect heterogeneity is crucial for improving how and to whom future interventions can be designed and delivered. In this article, we expanded on the works of van Breukelen and Candel\cite{van_breukelen_efficient_2015} and Moerbeek\cite{moerbeek_optimal_2020} to develop several optimal design procedures for obtaining the required cluster size and number of clusters that maximize statistical power to test for HTE in CRTs based on a pre-specified effect modifier, measured at either the individual level or cluster level, under a budget constraint (in other words, examining the cost-effectiveness of CRT designs for properly studying important treatment effect moderation). We further extended this optimal design procedure to allow for uncertainty in the assumed covariate-ICC and outcome-ICC values, to achieve design robustness to ICC value misspecification. Our new methodology is further illustrated using a recent CRT that published information on costs for sampling clusters and individuals, which assisted in the ascertainment of optimal design. As we elaborate in Section \ref{s:intro} and Table \ref{tab:LODlit}, existing optimal design methodology for CRTs had largely focused on maximizing power for testing the average treatment effect and had not yet considered treatment effect heterogeneity with respect to baseline effect modifiers or covariates, nor had they considered maximin designs that are based on two objectives (testing for the average treatment effect and pre-specified treatment effect heterogeneity). This paper fills those important methodological gaps. 

Of note, we have pursued the optimal design results with a quantitative endpoint analyzed by linear mixed models, whereby under this framework, the optimal design critically depends on the variance of the target estimator and is free of the effect size. This has been noted for optimal design results for assessing the average treatment effect,\cite{raudenbush_statistical_1997} and is also applicable when the interest lies in studying HTE (single-objective optimal design) and in studying both the average and heterogeneous treatment effect (multiple-objective optimal design). Furthermore, as was discussed in Section \ref{s:power}, it is important to notice that the total budget specified is often a key consideration to ensure a practical design where the testing objectives are properly powered; the optimal design under the budget constraints will then boil down to the optimal number of clusters ($n$) as optimal cluster size ($m$) then only depends on the cluster-to-individual cost ratio. If either the single- or multiple-objective maximin design provides insufficient power for the desired effect size(s), the total budget must be increased to provide a larger statistical power. Finally, we note that like van Breukelen et al,\cite{van_breukelen_efficient_2015} Liu et al,\cite{liu_optimal_2019} and Moerbeek,\cite{moerbeek_optimal_2020} we base our maximin procedure on a function of relative efficiency. It is also natural to consider a procedure that maximizes the minimum efficiency (equivalently, minimizing the maximum $\sigma^2_{\HTE}$ and thus maximizing the minimum power); the procedure in this case would always result in the LOD for the worst-case ICC combination scenario (or as close to the LOD as we may get in the chosen design space), which may potentially be a very different design than that identified under a RE-based maximin procedure. As noted in van Breukelen et al,\cite{van_breukelen_efficient_2015} however, an efficiency-based maximin procedure for studying the average treatment effect has the potential to be very inefficient if the true ICC values are very different than the worst-case scenario. Further investigations are necessary to elucidate the operating characteristics of such an efficiency-based maximin design when the interest lies in assessing treatment effect heterogeneity.

There are several limitations and possible future extensions of our current work. First, we only considered the case where the aim lies in testing treatment effect heterogeneity or moderation with respect to a univariate baseline covariate, which can be either binary or continuous. While this is a common scenario in studying confirmatory HTE and where sample size calculations at the design stage require relatively fewer parameters, it might be of interest to extend our framework to a joint test for HTE with respect to multiple or multivariate effect modifiers. Even for the single-objective design, this extension requires one to properly define the optimality criterion in terms of a variance-covariance matrix of the interaction parameter estimators; see Yang et al\cite{yang_sample_2020} for a characterization of the variance matrix expression (which we refer to as $\Sigma_{\HTE}$ in subsequent text) that extends $\sigma^2_{\HTE}$ with multiple covariates. For example, it would be worthwhile to identify locally optimal designs by minimizing either the trace of the variance-covariance matrix $\Sigma_{\HTE}$, the determinant of $\Sigma_{\HTE}$, or the minimum eigenvalues of $\Sigma_{\HTE}$; these three optimality criteria are akin to the A-optimality, D-optimality and E-optimality in the classic optimal design literature.\cite{fedorov_theory_2013} Extensions of any of these locally optimal designs to maximin designs open up new avenues for additional research. Second, as is conventional for planning CRTs, we have assumed a constant cluster size $m$, whereas in practice the cluster sizes may be variable due to non-informative drop-out or the fact that the source population is heterogeneous. While the extension of our optimal designs for assessing HTE and the compound objective is worthy of further investigation, Tong et al\cite{tong_accounting_2021} has recently pointed out that the ``correction factor'' for $\sigma^2_{\HTE}$ based on an individual-level effect modifier due to cluster size variation is almost equal to $1$ in a wide range of the parameter space. This suggests that our optimal design procedure for assessing HTE with an individual-level effect modifier would likely be robust under small to moderate degrees of cluster size variability. The correction factor for $\sigma_{\HTE}^2$ with a cluster-level effect modifier shares the same form with that derived earlier in van Breukelen et al,\cite{van_breukelen_relative_2007} though it can exceed $1$ and even be as large as $1.24$.\cite{tong_accounting_2021} We plan to conduct additional research to elucidate the impact of unequal cluster sizes for identifying the optimal designs based on the compound objective in Section \ref{s:mood}. Finally, we also assume the cluster-to-individual cost ratio does not vary by study arm. There are cases where the cost ratio will differ between treatment and control arms, as might also be the case in the K-DPP study. The heterogeneous cost ratio may therefore lead to a different locally optimal or maximin design, which will require further modifications of our procedure{; the same would be true if each study arm had different total budgets allocated to each or if there were heterogeneity in outcome variance between the arms (such as when a binary outcome is considered). Van Breukelen and Candel\cite{van_breukelen_maximin_2021} recently developed a single-objective maximin design for studying the average treatment effect by accommodating cost as well as variance } heterogeneity, and Moerbeek\cite{moerbeek_optimal_2020} recently considered the case where both cost ratios and cluster size vary between arms, wherein the optimal design is expressed as a ratio of sample sizes. We plan to carry out future work along these directions to extend our Proposition \ref{PROP:SOLOD} and Proposition \ref{PROP:MOLOD}, and to refine the associated operational details for achieving cost-effective study designs in broader settings with categorical outcomes and cost heterogeneity.

\section*{acknowledgements}
Research in this article was supported by a Patient-Centered Outcomes Research Institute Award\textsuperscript{\textregistered} (PCORI\textsuperscript{\textregistered} Award ME-2020C3-21072), and by CTSA Grant Number UL1 TR001863 from the National Center for Advancing Translational Science (NCATS), a component of the National Institutes of Health (NIH). The statements presented in this article are solely the responsibility of the authors and do not necessarily represent the views of PCORI\textsuperscript{\textregistered}, its Board of Governors or Methodology Committee, or the National Institutes of Health.

\section*{Data Availability Statement}
Data used in this article as an illustrative example are publicly available from the figshare database at \url{https://doi.org/10.6084/m9.figshare.5661610}.


\appendix
\section{Proof of Proposition \ref{PROP:SOLOD}}\label{soLODProof}
To find a closed-form solution for the LOD for assessing HTE, first recall HTE variance equation \eqref{eq:sHTE2}:
\begin{equation*}
    \sigma^2_{\HTE} \propto \frac{s(1-\rho_{y|x})}{B} \times
    \frac{(k+m)\left[1+(m-1)\rho_{y|x}\right]}{m\left[1+(m-2)\rho_{y|x} - (m-1)\rho_x\rho_{y|x}\right]}.
\end{equation*}

\noindent Taking the derivative with respect to $m$, we get:
\begin{align*}
    &\frac{\left\{\left[1+(m-1)\rho_{y|x}\right] + (k+m)\rho_{y|x}\right\}m\left[1+(m-2)\rho_{y|x} - (m-1)\rho_x\rho_{y|x}\right]}{\left\{m\left[1+(m-2)\rho_{y|x} - (m-1)\rho-x\rho_{y|x}\right]\right\}^2}\\
    &~~- \frac{(k+m)\left[1+(m-1)\rho_{y|x}\right]\left\{\left[1+(m-2)\rho_{y|x}-(m-1)\rho_x\rho_{y|x}\right] + m(\rho_{y|x}-\rho_x\rho_{y|x})\right\}}{\left\{m\left[1+(m-2)\rho_{y|x} - (m-1)\rho-x\rho_{y|x}\right]\right\}^2}.
\end{align*}

\noindent This can be simplified to:
\begin{equation}\label{appEq:derivativeHTE}
    (b_1 - ka_1)m^2 - ka_2 - ka_3,
\end{equation}
where
\begin{align*}
    a_1 &= \rho^2_{y|x}(1-\rho_x),\\
    a_2 &= 2\rho_{y|x}(1-\rho_{y|x})(1-\rho_x),\\
    a_3 &= (1-2\rho_{y|x}+\rho_x\rho_{y|x})(1-\rho_{y|x}),\\
    b_1 &= \rho_{y|x}(\rho_x - \rho_{y|x}).
\end{align*}

\noindent Setting \eqref{appEq:derivativeHTE} equal to 0, we can solve for $m$ through the quadratic equation:
\begin{equation*}
    m=\frac{ka_2 \pm \sqrt{k^2a^2_2 + 4ka_3(b_1 - ka_1)}}{2(b_1 - ka_1)},
\end{equation*}
which can be further simplified to:
\begin{equation*}
    m_{\opt}=\frac{(1-\rho_{y|x})(1-\rho_x) \pm \sqrt{\rho^{-1}_{y|x}k^{-1} (1-\rho_{y|x})(\rho_x - \rho_{y|x}) \left[1-(k+2)\rho_{y|x} + (k+1)\rho_x\rho_{y|x}\right]}}{k^{-1}(\rho_x - \rho_{y|x}) - \rho_{y|x}(1-\rho_x)}
\end{equation*}
focusing on the additive root as this is where solutions greater than 0 will occur:
\begin{equation}
    m_{\opt}=\frac{(1-\rho_{y|x})(1-\rho_x) + \sqrt{\rho^{-1}_{y|x}k^{-1} (1-\rho_{y|x})(\rho_x - \rho_{y|x}) \left[1-(k+2)\rho_{y|x} + (k+1)\rho_x\rho_{y|x}\right]}}{k^{-1}(\rho_x - \rho_{y|x}) - \rho_{y|x}(1-\rho_x)}
\end{equation}

\section{Proof of Proposition \ref{PROP:MOLOD}}\label{moLODProof}
To find a closed-form solution for the multiple-objective LOD, we recall the variance equation \eqref{eq:sHTE2}:
\begin{equation*}
    \sigma^2_{\HTE} \propto \frac{s(1-\rho_{y|x})}{B} \times
    \frac{(k+m)\left[1+(m-1)\rho_{y|x}\right]}{m\left[1+(m-2)\rho_{y|x} - (m-1)\rho_x\rho_{y|x}\right]}.
\end{equation*}
We can find a similar expression for $\sigma^2_{\ATE}$ by rearranging cost function \eqref{eq:budget} for n and substituting this into average treatment effect variance equation \eqref{eq:sATE}:
\begin{equation}
    \sigma^2_{\ATE} \propto \frac{(k+m)\left[1+(m-1)\rho_{y|x}\right]}{m}.
\end{equation}

\noindent Using these in multiple objective optimality criterion \eqref{eq:optCritRE}, we can simplify the criterion to:
\begin{equation*}
    \Theta(\zeta | \lambda ) \propto w_{\ATE} \times \frac{m}{(k+m)\left[1+(m-1)\rho_{y|x}\right]} + w_{\HTE} \times\frac{m\left[1+(m-2)\rho_{y|x} - (m-1)\rho_x\rho_{y|x}\right]}{(k+m)\left[1+(m-1)\rho_{y|x}\right]},
\end{equation*}
where $w_{\ATE} = \lambda\Theta_{\ATE}(\zeta^*_{\ATE})$ and $w_{\HTE} = (1-\lambda)\Theta_{\HTE}(\zeta^*_{\HTE})$.

Taking the derivative of the above expression with respect to $m$, we get:
\begin{equation}\label{appEq:derivMO}
    w_{\ATE}\times \frac{k(1-\rho_{y|x}) -m^2\rho_{y|x}}{\left\{(k+m)\left[1+(m-1)\rho_{y|x}\right]\right\}^2} + w_{\HTE}\times \frac{(ka_1-b_1)m^2 +ka_2m +ka_3}{\left\{(k+m)\left[1+(m-1)\rho_{y|x}\right]\right\}^2},
\end{equation}
where 
\begin{align*}
a_1&=\rho^2_{y|x}(1-\rho_x),\\
a_2&=2\rho_{y|x}(1-\rho_{y|x})(1-\rho_x),\\
a_3&=(1-2\rho_{y|x}+\rho_x\rho_{y|x})(1-\rho_{y|x}),\\
\text{and } b_1&=\rho_{y|x}(\rho_x-\rho_{y|x}).
\end{align*}

\noindent Setting this equal to $0$, we can simplify the left hand side and group terms by the degree of $m$:
\begin{equation}\label{appEq:derivMOEq0}
    \left\{w_{\HTE}(ka_1 - b_1) - w_{\ATE}\rho_{y|x}\right\}m^2 + w_{\HTE}ka_2m + \left[w_{\ATE}k(1-\rho_{y|x} + w_{\HTE}ka_3\right] = 0.
\end{equation}

Using the quadratic equation, the solution where $m>0$ is:
\begin{equation}
    m_{\opt} = \frac{-w_{\HTE}ka_2 - \sqrt{w_{\HTE}^2k^2a^2_2 - 4\left[w_{\HTE}(ka_1 - b_1) - w_{\ATE}\rho_{y|x}\right]\left[w_{\ATE}k(1-\rho_{y|x}) + w_{\HTE}ka_3\right]}}{2\left[w_{\HTE}(ka_1 - b_1) - w_{\ATE}\rho_{y|x}\right]}
\end{equation}

\noindent When $\rho_x=1$, $\sigma^2_{\HTE} = \sigma^2_{\ATE}$, making the optimality criterion RE of the average treatment effect and derivative \eqref{appEq:derivMO} simplifies such that we have:
$$\Theta_{\ATE}(\zeta^*_{\ATE})\times \frac{k(1-\rho_{y|x}) -m^2\rho_{y|x}}{\left\{(k+m)\left[1+(m-1)\rho_{y|x}\right]\right\}^2}.$$
Setting this equal to $0$ and solving for $m$ we get:
\begin{equation}\label{appEq:moLODrhox1}
    m_{\opt} = \sqrt{\frac{(1-\rho_{y|x})}{\rho_{y|x}}\times k}.
\end{equation}

\section{Supplementary Figures}\label{3dplots}

\begin{figure}
    \includegraphics[width=\textwidth]{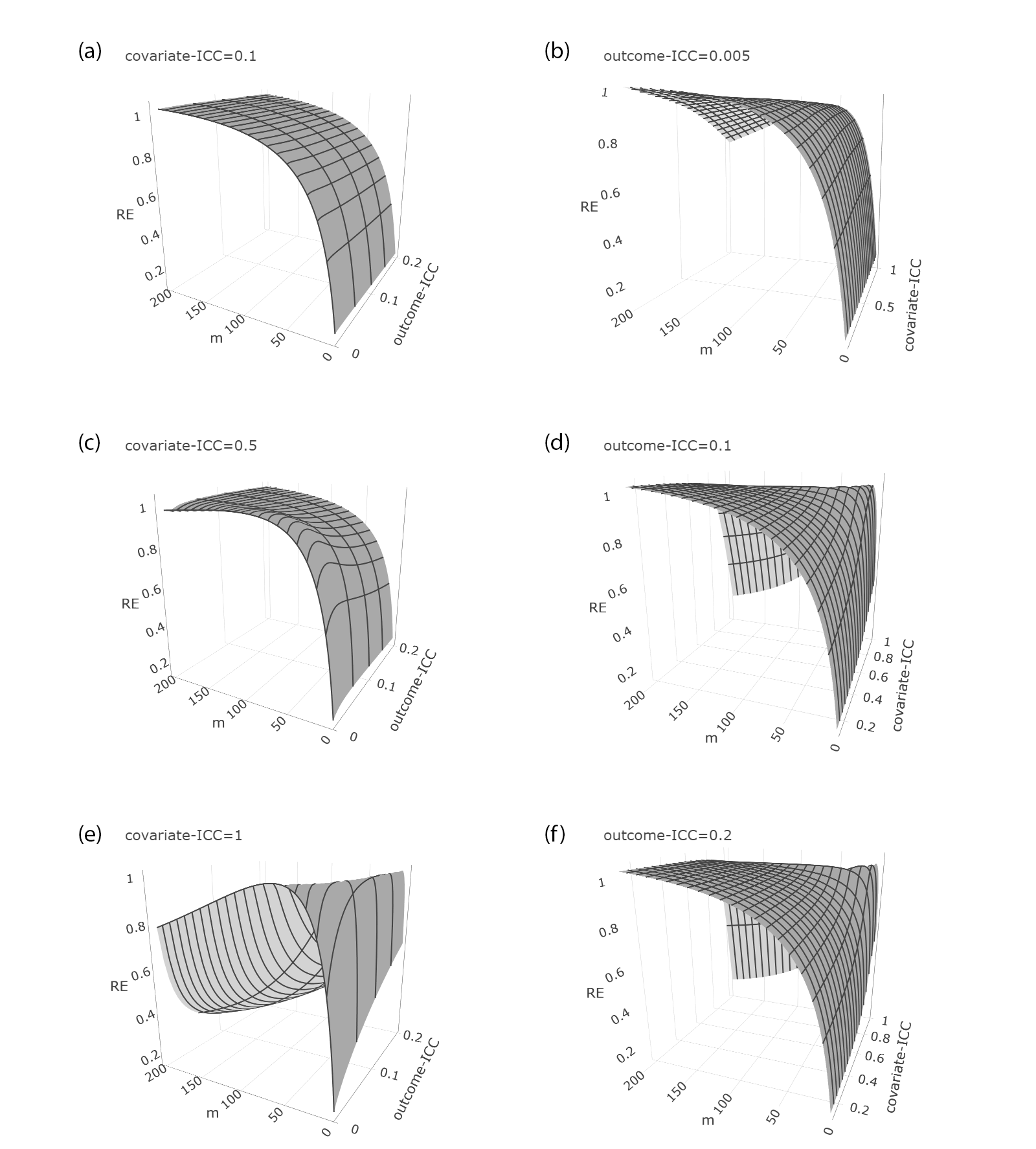}
    \caption{\label{fig:soMMD-3d}Three-dimensional plots of relative efficiencies (RE) of designs with cluster size $m$ versus their respective LODs for a cluster-to-individual cost ratio of $20$. Each surface represents the RE across continuous $m$ and either continuous $\rho_{y|x}$ for fixed $\rho_x \in \{0.1, 0.5, 1\}$ (Panels: a, c, e) or continuous $\rho_x$ for fixed $\rho_{y|x}\in \{0.005, 0.1, 0.2\}$ (Panels: b, d, f).}
\end{figure}

\begin{figure}
    \includegraphics[width=\textwidth]{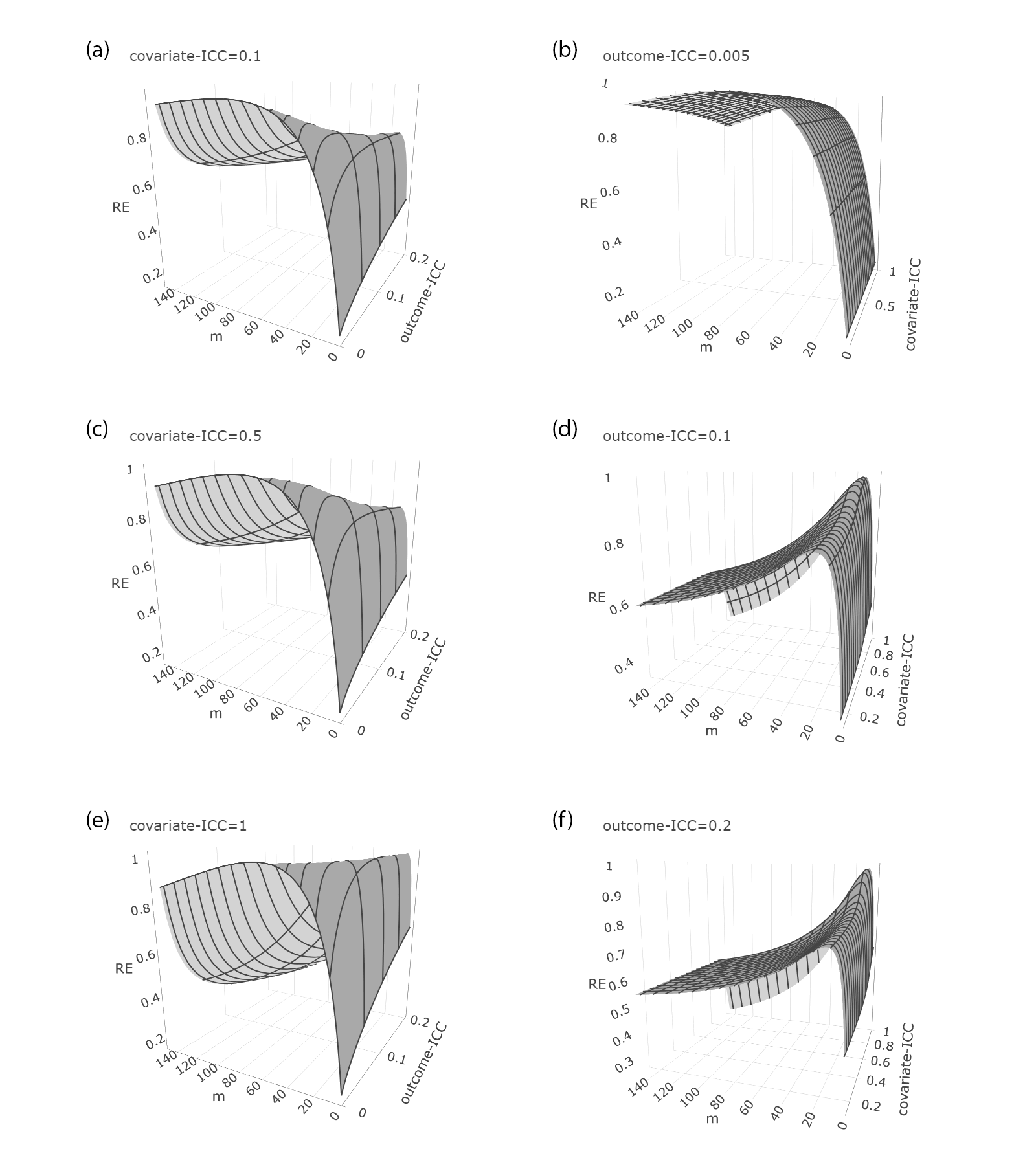}
    \caption{\label{fig:moMMD-3d}Three-dimensional plots of weighted relative efficiencies (RE) of designs with cluster size $m$ versus their respective LODs for a cluster-to-individual cost ratio of $20$ and a priority weight $\lambda=0.6$. Each surface represents the weighted RE across continuous $m$ and either continuous $\rho_{y|x}$ for fixed $\rho_x \in \{0.1, 0.5, 1\}$ (Panels: a, c, e) or continuous $\rho_x$ for fixed $\rho_{y|x}\in \{0.005, 0.1, 0.2\}$ (Panels: b, d, f).}
\end{figure}

\begin{figure}
\centering
[Figure C3 Here]
\includegraphics[width=0.95\textwidth]{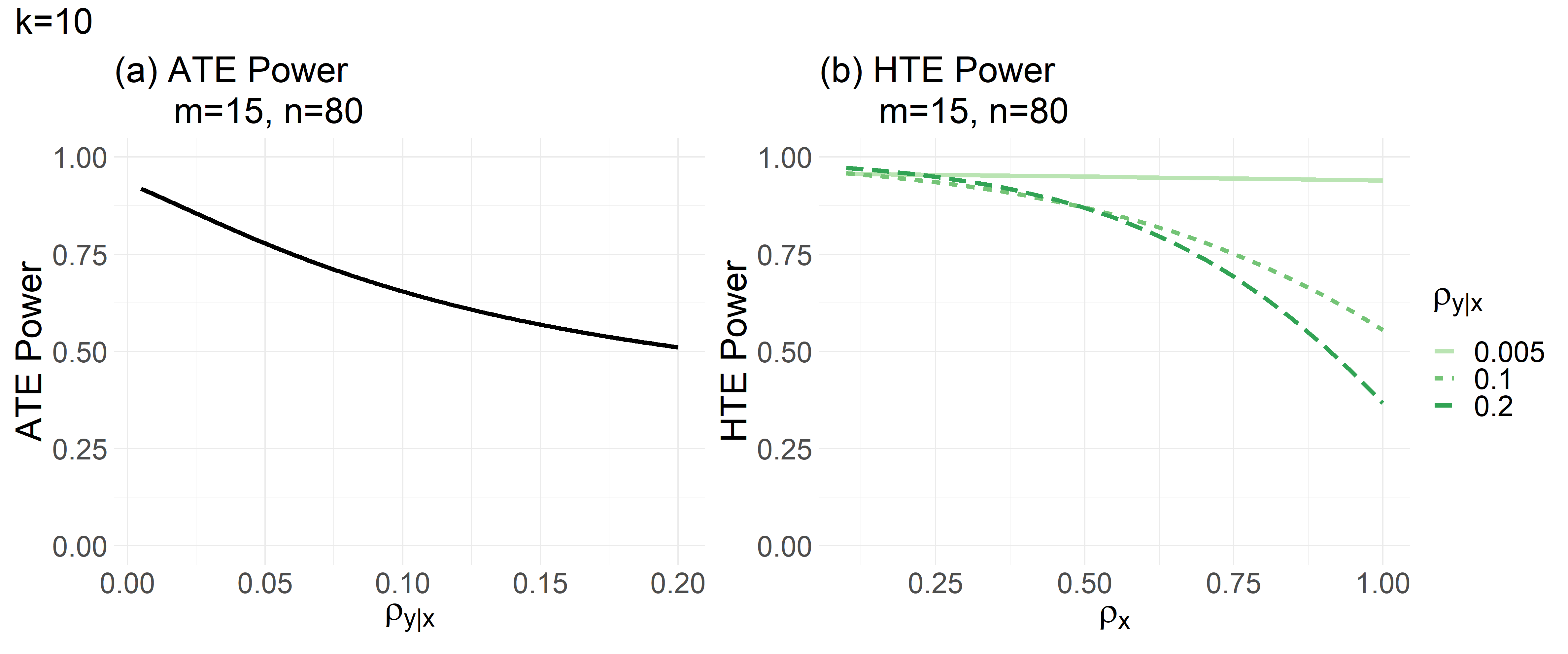}
\includegraphics[width=0.95\textwidth]{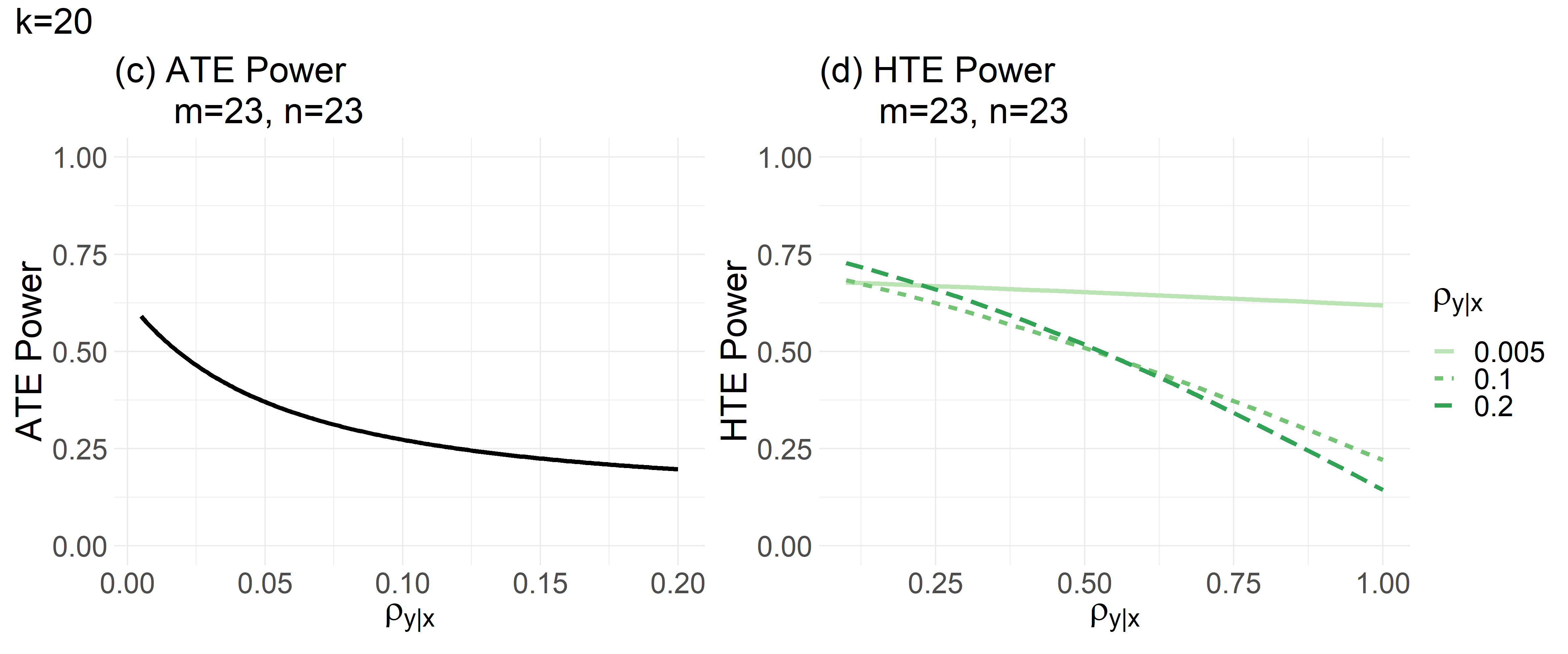}
\raggedleft\caption{\label{fig:moMMD-power}Heterogeneous (Panels: b,d) and average treatment effect (Panels: a,c) power curves for a standardized effect sizes of $0.2$ across $\rho_x$ and $\rho_{y|x}$ values for a cluster-individual cost ratio of $10$  (Panels: a,b) and  $20$ (Panels: c,d), evaluated at their respective maximin designs.}
\end{figure}
\renewcommand*{\thetable}{\arabic{table} }

\end{document}